\documentclass[useAMS,usenatbib,onecolumn,referee]{mn2e}

\usepackage{dcolumn}
\usepackage{graphicx}
\usepackage{url}
\usepackage{color}
\usepackage{pdflscape}

\newcommand{\cm}{cm$^{-1}$}

\newcommand{\X}{$X\,{}^1\Sigma^+$}
\newcommand{\A}{$A\,{}^1\Pi$}
\newcommand{\ap}{$a\,{}^3\Pi$}
\newcommand{\B}{$B\,{}^1\Sigma^+$}
\newcommand{\bp}{$b\,{}^3\Sigma^{+}$}

\newcommand{\allstates}{\X, \A, \B, \ap\ and \bp}

\newcommand{\ai}{\textit{ab initio}}

\newcommand{\eqref}[1]{(\ref{#1})}

\newcommand{\duo}{{\sc Duo}}
\newcommand{\Duo}{{\sc Duo}}

\newcommand{\xcross}{{\sc ExoCross}}

\usepackage{epstopdf}%This line makes .eps figures into .pdf - please comment out if not required.

\title[ExoMol XXXII: The spectrum of MgO]{ExoMol line lists XXXII: The rovibronic spectrum of MgO}

\date{\today}
\author[Li et al]{\large {Heng Ying Li, Jonathan Tennyson\thanks{Email: j.tennyson@ucl.ac.uk}  and Sergei N. Yurchenko}\\
Department of Physics and Astronomy, University College London, London WC1E 6BT, UK}
\date{Accepted XXXX. Received XXXX; in original form XXXX}

\pagerange{\pageref{firstpage}--\pageref{lastpage}} \pubyear{2018}

\begin{document}

\maketitle

\begin{abstract}

  Line lists for magnesium oxide are computed  and extensive
comparisons are made with existing
   experimental spectra. The LiTY line lists
  cover all ro-vibration transitions within the five lowest-lying
  electronic states ($X\,{}^1\Sigma^+$, $a\,{}^3\Pi$, $A\,{}^1\Pi$,
  $B\,{}^1\Sigma^+$ and $b\,{}^3\Sigma^{+}$) and five isotopologues:
  $^{24}$Mg$^{16}$O, $^{25}$Mg$^{16}$O, $^{26}$Mg$^{16}$O,
  $^{24}$Mg$^{17}$O, $^{24}$Mg$^{18}$O, $^{24}$Mg$^{17}$O and
  $^{24}$Mg$^{18}$O. The calculation use potential energy cures,
  spin-orbit and electronic angular momentum couplings curves
  determined by fitting to empirical energy levels; these levels are
  reproduced to within 0.01 \cm\ in most cases. Computed nuclear-motion
  wavefunctions are combined with {\it ab initio} dipole moment curves to
  give transition intensities and excited state radiative lifetimes
  which are compared with laboraroty measurements.  The
  $^{24}$Mg$^{16}$O line list comprises 186~842 ($J\le 320$)
  ro-vibronic states and 72~833~173 transitions with angular
  momenta, $J$, up to 300 and covering wavenumbers up to
  33~000 cm$^{-1}$ ($\lambda > 0.3$~$\mu$m). .The line
  lists are suitable for temperatures up to about 5000~K. They are
  relevant to astrophysical studies of exoplanet atmospheres, cool
  stars and brown dwarfs,  and are made available in electronic form at
  the CDS and ExoMol databases.

\end{abstract}

\label{firstpage}

\section{Introduction}

Magnesium oxide (MgO) is a metal oxide formed from the cosmically
abundant elements of magnesium and oxygen.
MgO is a major constituent
of chondritic meteorites \citep{10MaBeYa.MgO}.
It
was observed in the exosphere of Mercury by the first and
second flyby of the MESSENGER spacecraft
\citep{10KiPoVe.MgO,11SaKiMc.MgO}. The observed distribution suggests
 temperatures of tens of thousands Kelvin.
The quantity of MgO in the lunar surface was estimated
using X-ray fluorescence observations made by the Chandrayaan-1 X-ray
Spectrometer \citep{12WeKeSw.MgO}. The
strongest singlet band (\B\ -- \X) present in the lunar
exosphere were analysed during the 2009 Perseid  meteor shower
\citep{14BeChKl.MgO}.

MgO is also thought to be
a major component of interstellar dust
\citep{95YoGrxx.MgO,03NoKoUm.MgO,09Rietme.MgO}. Attempts to
detect MgO in the gas phase in the interstellar medium (ISM) have so far
proved inconclusive \citep{85TuStxx.MgO} or negative
\citep{98SaWhKa.MgO}, suggesting that Mg is heavily depleted onto
grains \citep{98SaWhKa.MgO}. The discovery of hot, rocky exoplanets
has led to suggestions that the many compounds usually
in the condensed phase, and in particular
MgO \citep{13CoSmEg.MgO}, should be present in significant
concentrations in the gas phases \citep{12ScLoFe.exo,jt731}. The detection of us such species
would require lists of key spectroscopic transitions and here we
provide line lists for MgO isotopologues which should be valid
over an extended range of temperatures

MgO has a very characteristic electronic spectrum which should be
amenable to observations. In particular,
the two singlet systems
(\B\ -- \A\  and \B\ -- \X), called the
red and green band systems respectively, have long been studied
in the laboratory \citep{32Maxxxx.MgO,49LaUhxx.MgO}.
Experimental studies of MgO spectra  initially used
grating spectrographs and more recently
laser spectroscopy  which allowed
experiments to be performed on the low lying states as well as some higher
states such as the $E\,{}^1\Sigma^+$ and $F\,{}^1\Pi$  states \citep{03BeBuVa.MgO,04WaVaBe.MgO}.
A full survey of high resolution spectroscopic studies of MgO is given
in the next section.

The most recent and comprehensive \ai\ study of MgO is by
\citet{17BaScxx.MgO}, who presented a high level study for a large
number of electronic states of this molecule using the
SA-CASSCF/IC-MRCI/aug-cc-pV5Z level of theory. They reported high
level \ai\ potential energy and (transition) dipole moment curves as
well as lifetimes for a number of electronic bands. We have adopted
these \ai\ curves for the present study.  Another recent and accurate
\ai\ study of MgO, also relevant to our work, is by
\citet{10MaBeYa.MgO}, who used the MRCI/cc-pV5Z method to compute
potential energy, spin-orbit and transition dipole moment curves for
the valence and valence-Rydberg electronic states of MgO. Earlier,
but also comprehensive, \ai\ studies of MgO were performed by
\citet{89ThKlP1.MgO} and \citet{89ThKlPe.MgO} using the MRDCI method
for various states.  None of these works reported the electronic angular
momenta of MgO.

Limited work have been performed on the construction of MgO line
lists.  While studies have reduced the observed spectra into
spectroscopic constants
\citep{87PyDiMu.MgO,89ThKlP1.MgO,02DaGrAb.MgO}, the only sourcesof
astronomical transition data is the JPL list of 44 pure rotational
transitions associated with the $v=0$ and $v=1$ vibrational states of
the \X\ electronic ground state \citep{jpl}. These data are designed
for detecting MgO in the interstellar medium but do not provided the
information required to study the spectra of hotter sources.

The ExoMol project
aims to provide a catalogue of spectroscopic transitions for molecular
species which may be present in the atmospheres of the exoplanets,
brown dwarfs and cool stars \citep{jt528}.  The ExoMol database
\citep{jt631} currently contains 52 molecules ranging from diatomic to
polyatomic molecules and ions \citep{jt731}.  However, there are only
a handful of metal oxides are available, namely CaO \citep{jt618}, VO
\citep{jt644}, AlO \citep{jt598} and SiO \citep{jt563}; an updated line
list for TiO has just been completed \citep{jt760}.
Such line lists are also important for analysing and modelling spectra
in laser induced plasmas \citep{12WoPaHo.TiO,14DeDeDe,15PaWoSu.Tio}.

Here we present new extensive line lists for MgO covering the
spectroscopy of its five lowest electronic states, \allstates. The
line list are computed using nuclear-motion program \Duo\ \citep{Duo} using a
combination of empirical and \ai\ potential energy curves (PECs),
spin-orbit curves (SOCs) and electronic angular momenum curves (EAMCs)
in conjunction with high level \ai\ (transition) dipole moment curves
(T)DMCs. The \ai\ curves were taken from the literature, where
available, or were computed as part of this work.  Given the
astronomical interest in MgO, the line lists are produced to cover an
extensive energy and spectroscopic range, up to 37~250~\cm\ and should
be applicable for temperatures up to at least 5000 K.  Line lists are
generated for the three stable isotopes of Mg and three stable
isotopes of O.

\section{Method}

\subsection{Experimental data}

Experimental transitions frequencies for $^{24}$Mg$^{16}$O involving
a total nine different electronic states were collected from various papers which are
summarised in Table \ref{t:source}. The MARVEL technique (Measured Active
Rotational-Vibrational Energy Levels) \citep{jt412,12FuCsa,jt750} was used to
determine empirical energy levels. A total of 2457 transitions
were collected from the listed sources and validated yielding
757   distinct empirical energy levels.
The MARVEL transitions and energy files are given as part of the supplementary
data. The MARVEL energy levels together with the experimental
transitions were used as \Duo\ inputs to refine our spectroscopic model (PECs,
SOCs and EAMCs).

\begin{table}
\caption{Experimental sources and maximal values of $J$, $v$ and energy term values (\cm) of high resolution spectroscopic data of MgO used in this work.  }
\begin{tabular}{llccccrrrr}
\hline\hline
Source&Method&State$'$&State$''$&$v'$&$v''$&$J'$&$J''$&$\tilde{E}'_{\rm max}$&$\tilde{E}''{\rm max}$\\
\hline
\citet{06KaKaxx.MgO}&Microwave&{$a\,{}^3\Pi$}&$a\,{}^3\Pi$&0&0&12&7&7&13\\
&&$X\,{}^1\Sigma^+$&$X\,{}^1\Sigma^+$&1&1&12&7&7&14\\
&&&&2&2&12&7&&\\
&&&&3&3&12&7&&\\
\hline
\citet{62BrBeTr.MgO}&Grating&$C\,{}^1\Sigma^-$&$A\,{}^1\Pi$&0&0&80&4&26414&26544\\
&&&&1&1&65&21&&\\
\hline
\citet{65TrEwxx.MgO}&Grating&$C\,{}^1\Sigma^-$&$A\,{}^1\Pi$&1&1&60&9&26387&26769\\
&&$D\,{}^{1}\Delta$&$A\,{}^1\Pi$&0&0&61&1&&\\
&&&&1&1&43&4&&\\
\hline
\citet{73Singhx.MgO}&Ebert&$G\,{}^1\Pi$&$A\,{}^1\Pi$&0&1&40&14&36388&39870\\
&&$G\,{}^1\Pi$&$X\,{}^1\Sigma^+$&0&0&38&16&&\\
&&&&0&1&40&16&&\\
\hline
\citet{73Singhx.MgO}&Ebert&$F\,{}^1\Pi$&$X\,{}^1\Sigma^+$&0&0&83&14&37686&37896\\
\hline
\citet{76AnBoPe.MgO}&Grating&$E\,{}^1\Sigma^+$&$A\,{}^1\Pi$&0&0&56&36&34304&34173\\
\hline
\citet{84AzDyGe.MgO}&LIF&$B\,{}^1\Sigma^+$&$X\,{}^1\Sigma^+$&0&0&40&0&19971&20062\\
&&&&1&1&56&0&&\\
\hline
\citet{91CiHeBl.MgO}&DLFR&$X\,{}^1\Sigma^+$&$X\,{}^1\Sigma^+$&1&0&32&6&766&792\\
&&&&2&1&28&7&&\\
\hline
\citet{91IpCrFi.MgO}&LIF&$B\,{}^1\Sigma^+$&{$a\,{}^3\Pi$}&0&0&48&4&16735&27256\\
&&&&1&1&48&6&&\\
&&$D\,{}^{1}\Delta$&{$a\,{}^3\Pi$}&0&0&33&1&&\\
&&&&1&1&17&1&&\\
\hline
\citet{94KaHiTa.MgO}&FTS&$A\,{}^1\Pi$&$X\,{}^1\Sigma^+$&1&0&56&0&2644&5452\\
&&&&2&0&67&3&&\\
&&&&3&0&58&3&&\\
&&&&2&1&31&11&&\\
&&&&0&1&37&8&&\\
\hline\hline
\end{tabular}\label{t:source}
\noindent
\mbox{}\\
{\flushleft
Microwave = Microwave spectroscopy\\
Grating = Grating spectrograph\\
Ebert = Ebert spectrograph\\
DLFR = Diode-laser flame spectrometer\\
LIF = Laser-induced fluorescence spectroscopy\\
FTS = Fourier Transform Spectrometer\\
}
\end{table}

\subsection{Spectroscopic model}

As with other metal oxides studied within the ExoMol projects, the
procedure used here to calculate the line list for MgO is to refine
the \ai\ PECs, SOCs and EAMCs using available experimental data
\citep{jt511}.

\begin{figure}
\includegraphics[width=0.8\textwidth]{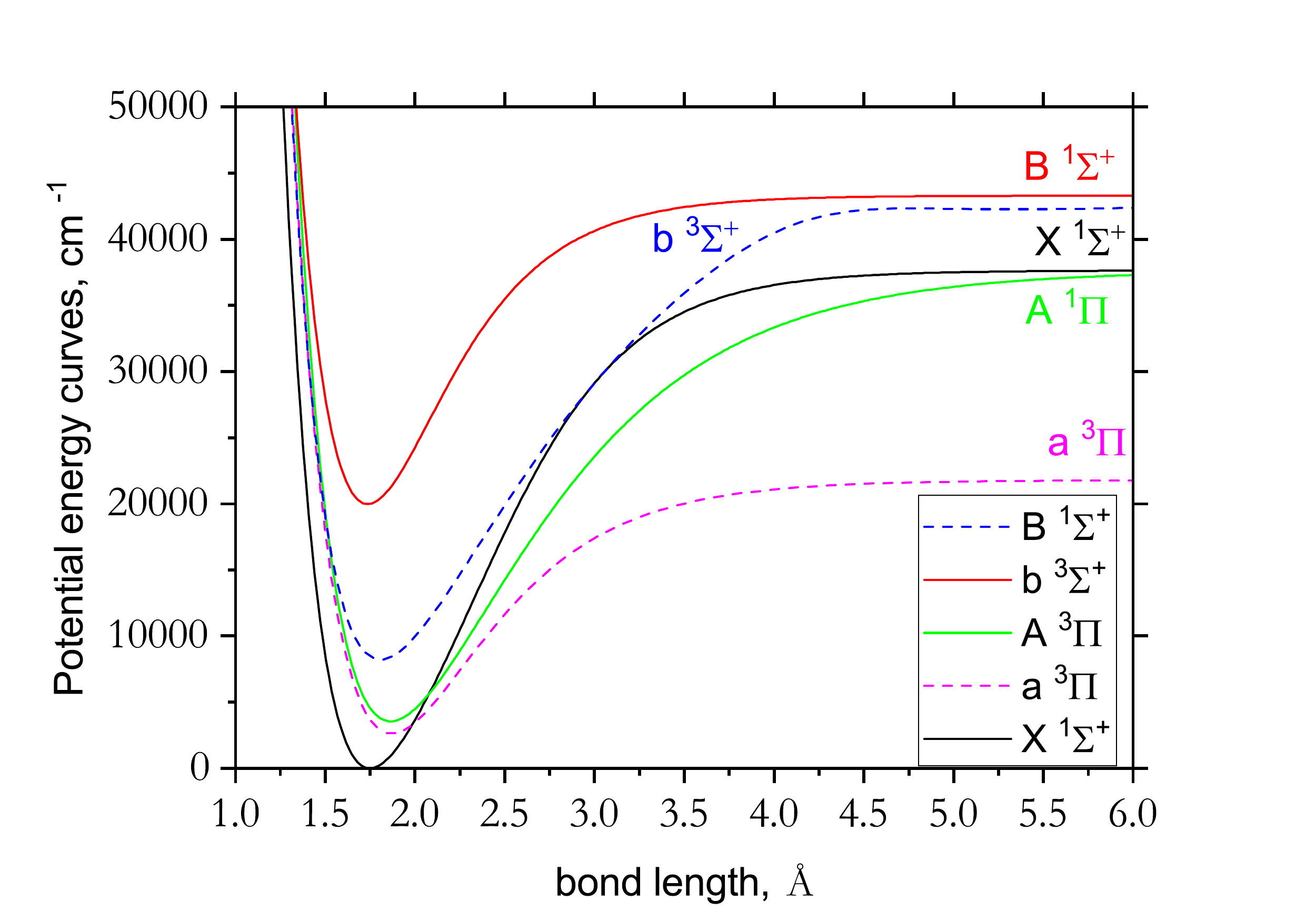}
\caption{Refined potential energy curves (PECs) of MgO used to produce the line list as part of the final spectroscopic model.}
\label{fig:PEC}
\end{figure}

\begin{figure}
\includegraphics[width=0.8\textwidth]{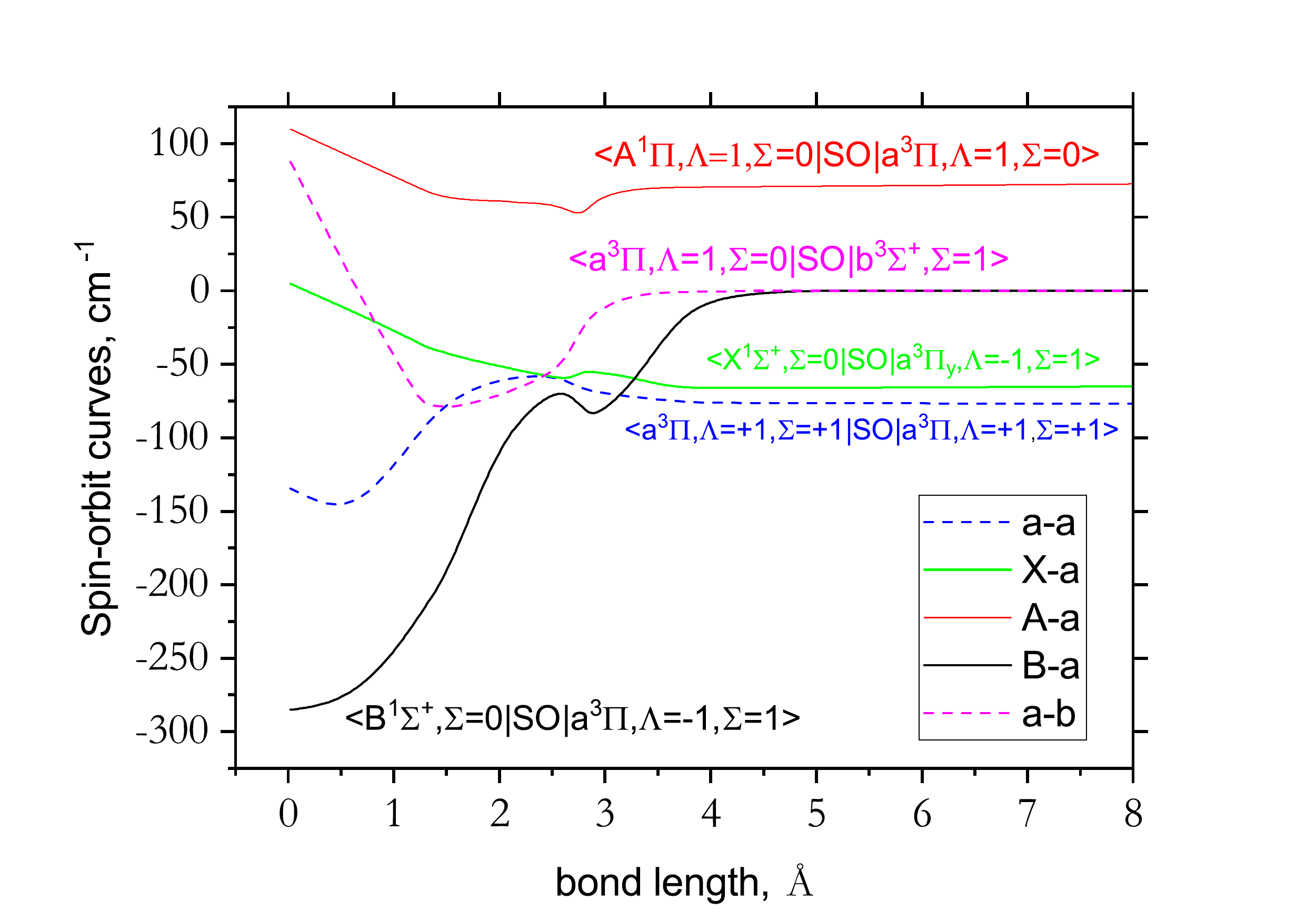}
\caption{Empirical spin-orbit (SO) couplings (\ai\ values morphed by \Duo) in the spherical representation. }
\label{fig:SO}
\end{figure}

\begin{figure}
\includegraphics[width=0.8\textwidth]{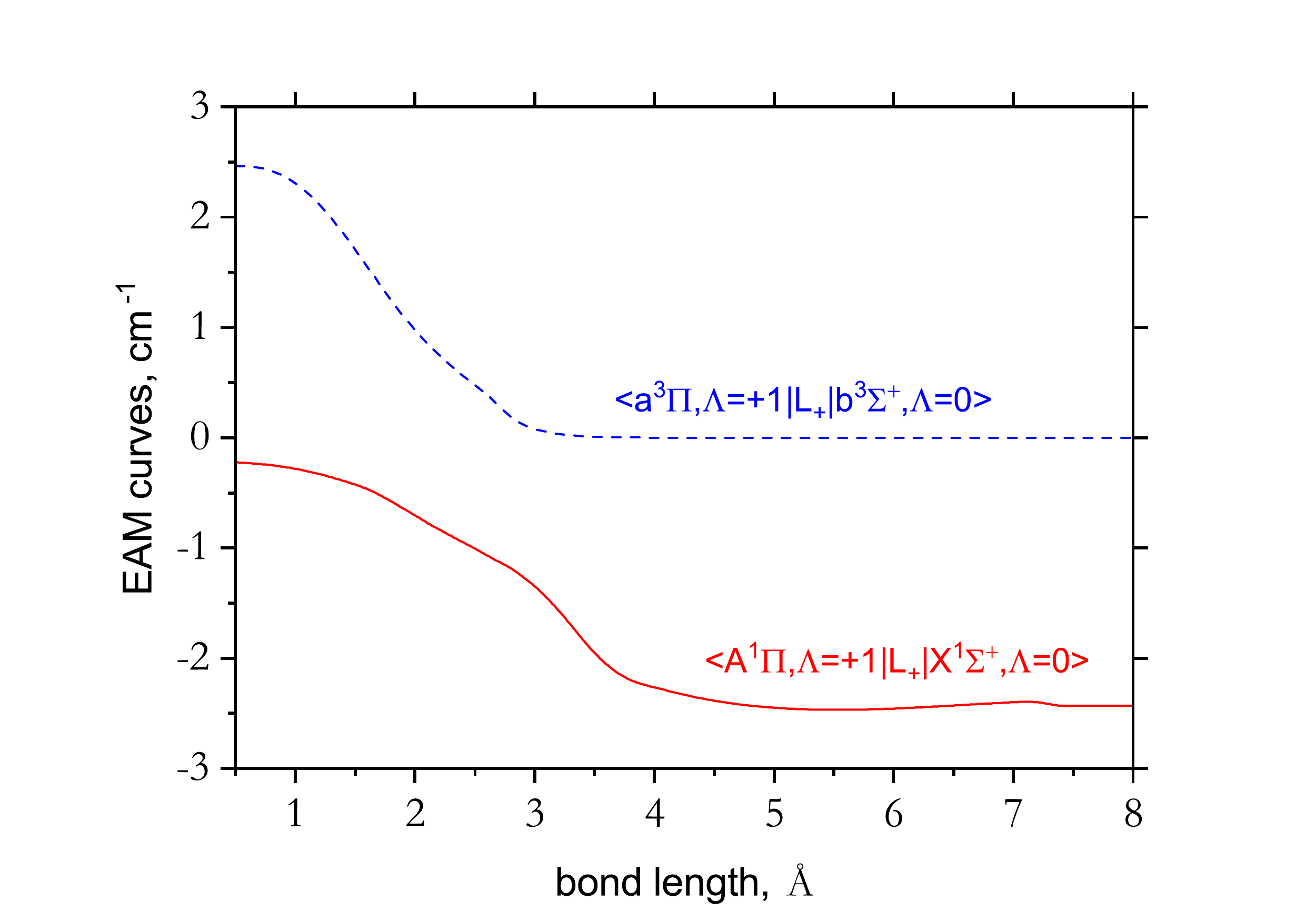}
\caption{Empirical Electronic Angular Momentum (\ai\ values morphed by \Duo), in the spherical representation.}
\label{fig:EMAC}
\end{figure}

Our model of MgO consists of five PECs, \allstates\ which are
shown in Figure~\ref{fig:PEC}. They are augmented by five SOCs,
$X$--$a$, $a$--$a$, $A$--$a$, $B$--$a$ and $a$--$b$, which are shown in Figure
\ref{fig:SO}, and two EAMCs $X$--$A$ and $a$--$b$, the $x$ components of which
are shown in Figure~\ref{fig:EMAC}. The initial, \ai, PECs  were taken from
\citet{17BaScxx.MgO}.  The SOCs and  EAMCs   were obtained \ai\ using the MOLPRO
  electronic structure package \citep{molpro} at the  multi-reference
configuration interaction (MRCI)  level of theory in conjunction with the
aug-cc-pwCVQZ basis sets with relativistic, core-correlation effects and
Davidson correction, as part of this study. The active space is given by
(8,4,4,0). The states-averaged CASSCF (SA-CASSCF) comprised (3,2,1,1)
states of symmetries $\Sigma^+$, $\Pi$, $\Delta$ and $\Sigma^-$, respectively.
All electron were correlated. The relativistic
correction was estimated using the Douglas-Kroll method.

%SOCs and EMACs involving the $B$ state have been ignored in this model as the $B$ state is high up in energy, the energy separation between the $B$ state and the other three states is very large. Therefore the $B$ state does not contribute much to SOCs and EMACs.

The nuclear motion program \Duo\ \citep{jt609} is used to solve the
fully-coupled Schr\"{o}dinger equation for the five lowest bound
electronic states of MgO.  \Duo\ calculations used a grid-based sinc
DVR basis of 501 points spanning 0.02 to 8 \AA\ to obtain vibrational
basis functions by solving five independent $J=0$ problems for the
corresponding electronic state.  The final vibrational basis set
comprised 100, 62, 100, 57, and 70 lowest vibrational eigenfunctions
of the $X$, $a$, $A$, $B$ and $b$ states, respectively. These numbers
were chosen to include all vibrational states below the corresponding
dissociation limits (see below). A detailed description of the
methodology is given by \citet{jt609} and \citet{jt632}.

%Duo calculates accurate empirical line list for diatomic molecules.

In order to facilitate the refinement of the experimental model, the
four lowest PECs ($X$, $a$ $A$, and $B$) were represented using an
extended Morse oscillator (EMO) \citep{EMO} potentials with the form:
\begin{equation}
\label{e:V}
	V(R) =V_{\rm e}+(A_{\rm e}-V_{\rm e})\left[1-\exp\left(-\sum_{k=0}^{N}B_k\xi^k(r-r_{\rm e})\right)\right]^2,
\end{equation}
where $D_{\rm e} = A_{\rm e}-V_{\rm e}$ is the dissociation energy,
$r_{\rm e}$ is an equilibrium distance of the PEC, and $\xi$ is the
\v{S}urkus variable \citep{84SuRaBo.method}. The \v{S}urkus variable
is defined as:
\begin{equation}
\xi= \frac{r^p-r^p_{\rm ref}}{r^p+r^p_{\rm ref}},
\end{equation}
where $p$ is a non-zero real number parameter, $r_{\rm ref}$ is a
reference position and $r_{\rm ref} = r_{\rm e}$ in this case. The $b$
state \ai\ PEC was not used in the refining procedure and therefore
did not need to be represented analytically. The \ai\ points values
were interpolated on the \duo\ grid using the cubic splines
\citet{Duo}.

The lowest dissociation limit from MgO is to the \ap\ asymptote
(Mg($^1S$) and O($^3P$)), which \citet{17BaScxx.MgO} estimated as lying
at 2.7$\pm 0.1$~eV. Their adiabatic dissociation energy for the \X\
state is 4.65~eV, which is similar to the value obtained using the
atomic asymptotic separation to O($^1D$), 4.67~eV. This is also the
asymptote as for the \A\ state. Adding the atomic separation of
O($^3P$) and Mg($^3P$) to this limit gives the asymptote for the \B\ state
of 5.4~eV. These asymptotes were adopted for the values of $A_{\rm e}$
in Eq.~\eqref{e:V} used in this work.

We did not represent the \ai\ SOCs and EAMCs analytically directly.
Instead, in order to allow for their refinement, the morphing
procedure \citep{99MeHuxx.methods, 99SkPeBo.methods, jt589} was used.
According to this procedure, as implemented in \Duo, the \ai\ SOCs and
EAMCs are multiplied by a morphing function $F(z)$ represented by
following expansion:
\begin{equation}
\label{e:F(z)}
	F(z)=\sum_{k=0}^{N}{B_k z^k(1-z)+z B_\infty},
\end{equation}
where $z$ is either taken as a damped-coordinate (for the $a$--$S$ SOC) given by:
\begin{equation}\label{e:damp}
z = (r-r_{\rm ref})\, e^{-\beta_2 (r-r_{\rm ref})^2-\beta_4 (r - r_{\rm ref})^4},
\end{equation}
see also \citet{jt703} and \citet{jt711}, or as the \v{S}urkus variable $z=\xi$ (all other SOCs and EAMCs).
Here $r_{\rm ref}$ is a reference position equal to $r_{\rm e}$ by default and $\beta_2$ and $\beta_4$ are damping factors.
When used with morphing, the parameter $B_{\infty}$ is usually fixed to 1.

 EAMCs give rise to the
$\Lambda$ doubling (parity doubling) for states with $\Lambda$
(projection of the electronic angular momentum) greater than 0.

The PECs, SOCs and EAMCs, representing the final, refined
spectroscopic model are shown in Figs.~\ref{fig:PEC}--\ref{fig:EMAC}
and given as part of the supplementary material inside the \duo\ input file.

%All \ai\ curves are provided by S. Yurchenko (ExoMol).

%The refined PECs are presented in Figure \ref{fig:PEC}.

\begin{figure}
\includegraphics[width=0.45\textwidth]{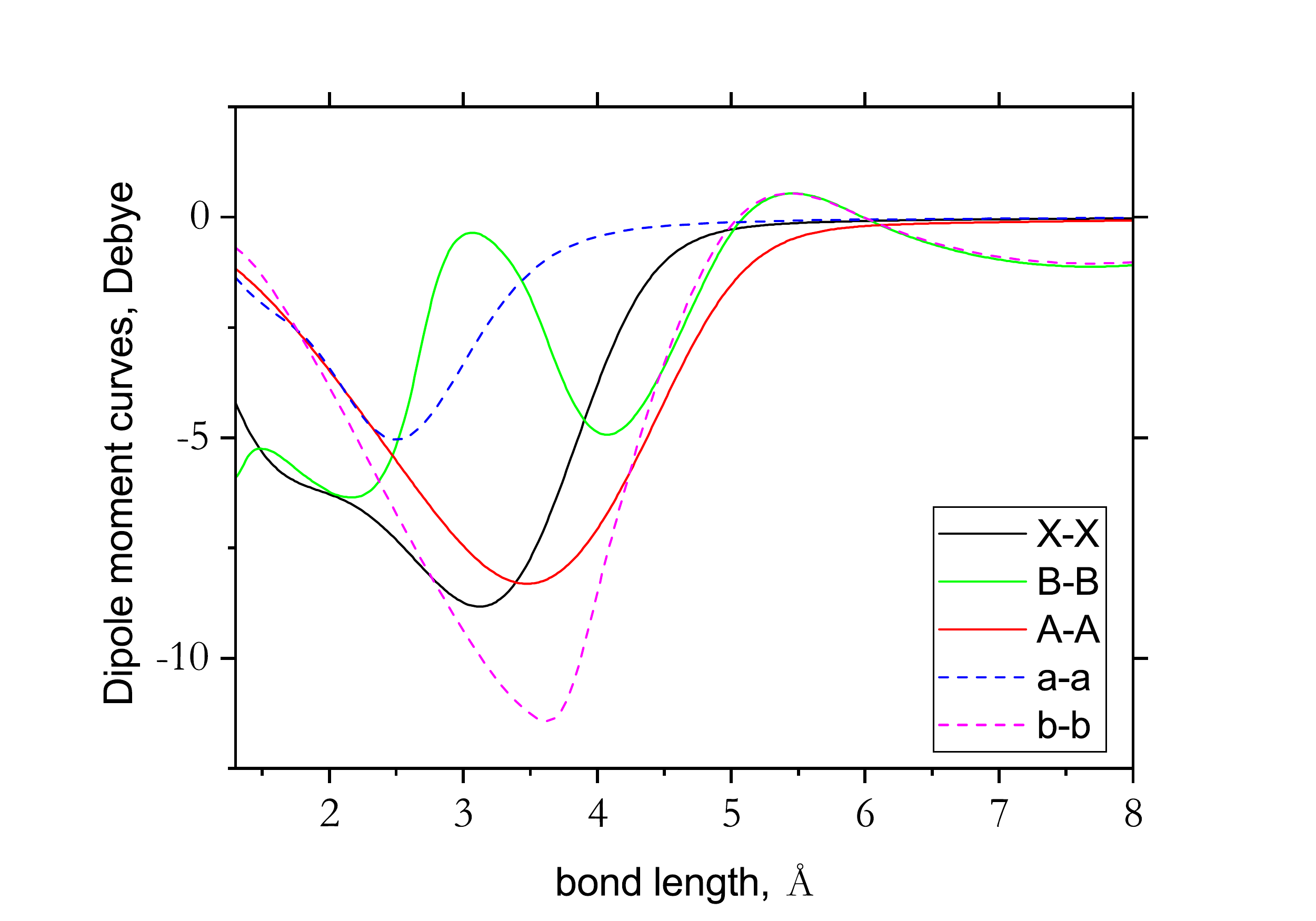}
\includegraphics[width=0.45\textwidth]{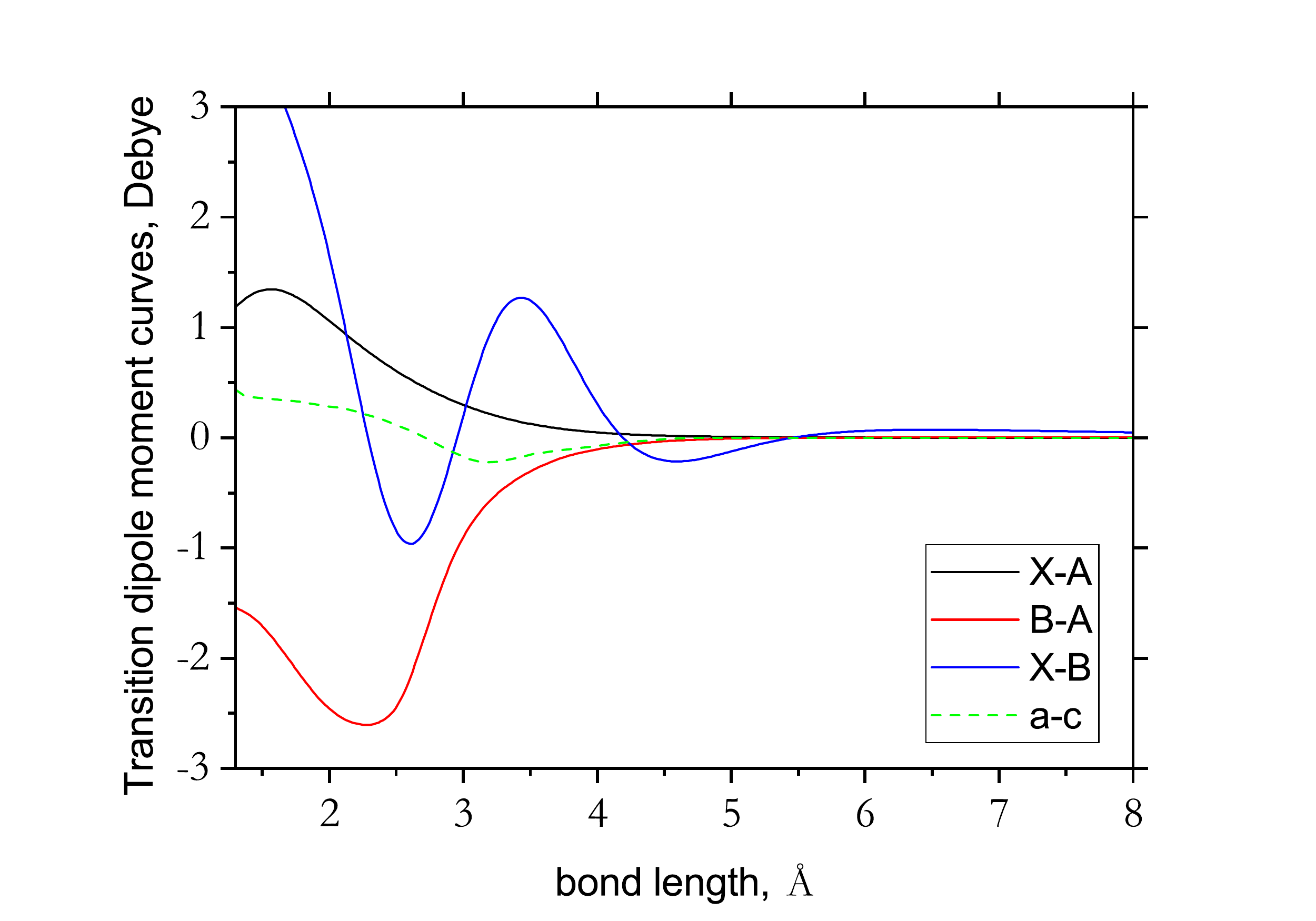}
\caption{Dipole moment curves and transition dipole moments due to \citet{17BaScxx.MgO} ($X-X$, $A-A$, $a-a$, $B-B$, $A-X$, $B-X$) and \citet{10MaBeYa.MgO} ($A-X$) presented as a function of the nuclear separation. }
\label{f:DMC}
\end{figure}

MgO is an ionic system. This leads to large permanent dipole moments
which vary strongly with $r$ and, because the degree of charge
separation changes strongly between excited states, large
transition dipoles. Similar behaviour was observed in the previous ExoMol
study on CaO \citep{jt618}.
The diagonal DMCs ($X-X$, $a-a$, $A-A$, $b-b$)  as well as the transition DMC
for $B-A$
are taken from the \ai\ work of \citet{17BaScxx.MgO} who used an MRCI
calculation with an
aug-cc-pV5Z. All other TDMCs ($A-X$, $B-X$, $a-c$) were computed \ai\ as part of
the present work.   The corresponding electronic transition dipole moments are
shown in Fig.~\ref{f:DMC}.
All (T)DMCs except for $B-B$, $B-A$, $b-b$    were represented analytically
using the damped-$z$ expansion in Eq.~\eqref{e:F(z)}.  This was done in order to
reduce the numerical noise in the calculated intensities for
high overtones, see recommendations by \citet{16MeMeSt}. The corresponding
expansion parameters as well as their grid
representations can be found in the \duo\ input files provided as supplementary
data. \citet{87BuHeHe.MgO} reported experimental electric dipole moments for the
\X\ and \A\ states ($v=0$) using Stark quantum-beat spectroscopy, 6.2(6) D and
5.94(24)~D, respectively. Our values of the vibrationally ($v=0$) averaged
dipole moments based on the \ai\ dipole moment of \citet{17BaScxx.MgO} are
5.99~D  and 5.68~D, respectively, which agree well with these experimental value.
The dipole moment curves (DMCs)  are shown in Figure
\ref{f:DMC}.

%The $A-X$ transition dipole moment curve is taken from
%\citet{10MaBeYa.MgO}  who performed an MRCI and cc-pV5Z basis
%set.

%B-B, B-A, b-b, A-A (analytical), а-а(analytical) is from 17Ba
%A-X, Б-Х, a-c are from this work

%X-X is from 17Ba, but a wrong factor used as used in line list calculations.

%It should be noted that the turning point of the \ai\ $B-X$ dipole moment of \citet{17BaScxx.MgO} at about 2.8 \AA\ might be an
%artifact of their calculations and it is possible that
% there should be a sign change in the dipole.

\Duo\ input files which fully specify our final spectroscopic model are given
as part of the supplementary data.

\section{Results}

\section{Accuracy of the fits}

\begin{figure}
\includegraphics[width=0.8\textwidth]{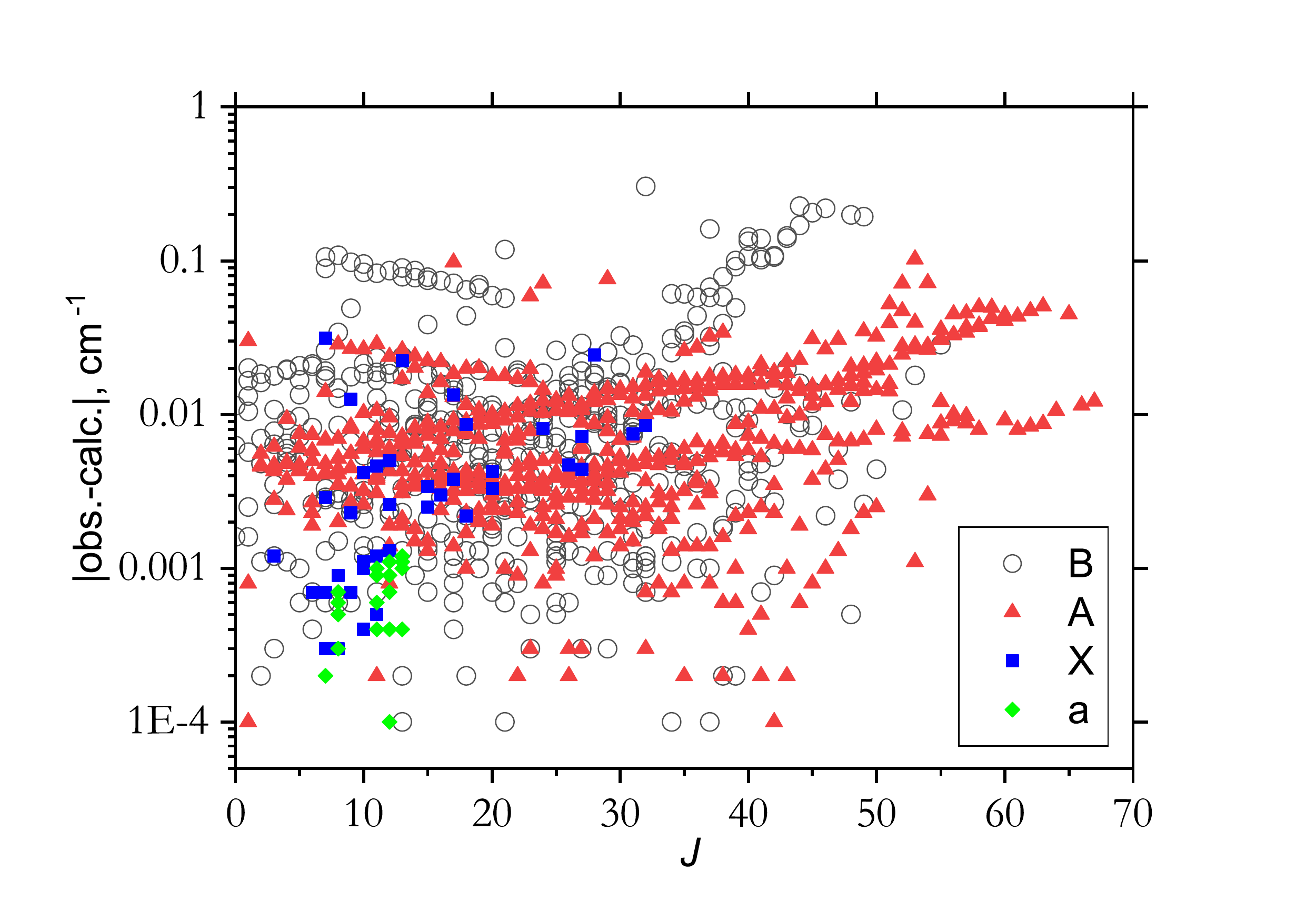}
\caption{The Obs. - Calc. residuals on different states as a function of the wavenumber }
\label{fig:error}
\end{figure}

The \ai\ PECs ans SOCs were refined by fitting to the 2457 experimental
transition frequencies (see Table~\ref{t:source}) augmented with 756 MARVEL term
values for $^{24}$Mg$^{16}$O. The data used in the fit are provided as part of the spectrospic model.
The root mean square (RMS) errors as observed minus calculated (Obs. $-$ Calc.)
residues are 0.029 \cm\ for all transition wavenumbers, or 0.009~\cm, 0.002~\cm,
0.016~\cm\ and 0.04~\cm\ for the $X$, $a$, $A$ and $B$ states, respectively. The
MARVEL term values are reproduced with an RMS error of 0.05~\cm.
Figure \ref{fig:error} gives an
overview of the Obs. $-$ Calc. residues with more detailed results given in
Tables~\ref{error_X},  \ref{t:error:A:B} and \ref{H2D0}.

Line lists for the other isotopologues of MgO were generated using the
curves with no allowance for any failure of the Born-Oppenheimer
approximation. This means that while we expect these line lists to
still be accurate, some loss of accuracy is to be expected.  The pure
rotational frequencies of $^{25}$Mg$^{16}$O and $^{26}$Mg$^{16}$O do
show excellent agreement with experiment, see Table~\ref{t:microwave}.

\begin{table}
\centering
\caption{Selections of the Obs.$-$Calc. residuals, in \cm, for the microwave transition frequencies ($X$, $v'=v"$, $R(J)$) of the three isotopologues of Mg$^{16}$O. The Obs. term values are from \protect\cite{86ToHoxx.MgO}.}
\label{t:microwave}
\begin{tabular}{rrrrrrrrrrrrrrrrrrrrr}
\hline
%\multicolumn{4}{c}{\X, $v=0$}   \\
\hline
Iso &$J$&$v$&Obs&Calc&Obs.-Calc. \\
\hline
$^{26}$MgO    &         1 &         0 &    2.217914 &        2.217857 &   0.000057  \\
$^{26}$MgO    &         2 &         0 &    3.326801 &        3.326716 &   0.000085  \\
$^{26}$MgO    &         2 &         1 &    3.296283 &        3.296159 &   0.000124  \\
$^{26}$MgO    &         7 &         0 &    8.869430 &        8.869204 &   0.000226  \\
$^{25}$MgO    &         2 &         0 &    3.377244 &        3.377161 &   0.000083  \\
$^{24}$MgO    &         1 &         0 &    2.288127 &        2.288066 &   0.000061  \\
$^{24}$MgO    &         1 &         1 &    2.266805 &        2.266722 &   0.000083  \\
$^{24}$MgO    &         2 &         0 &    3.432117 &        3.432026 &   0.000091  \\
$^{24}$MgO    &         2 &         1 &    3.400133 &        3.400008 &   0.000125  \\
$^{24}$MgO    &         2 &         2 &    3.368112 &        3.368431 &   -0.000319 \\
$^{24}$MgO    &         7 &         0 &    9.150138 &        9.149896 &   0.000242  \\
$^{24}$MgO    &         7 &         1 &    9.064847 &        9.064513 &   0.000334  \\
$^{24}$MgO    &         7 &         2 &    8.979447 &        8.980303 &   -0.000856 \\
\hline\hline
\end{tabular}

\end{table}

\begin{table}
\centering
\caption{Selections of the Obs.$-$Calc. residuals, in \cm, for the $X$ state and $a$ states  ($v=0$) for the refined model of $^{24}$Mg$^{16}$O. The Obs. term values are represented by our MARVEL values.}
\label{error_X}
\begin{tabular}{rrrrrrrrrrrrrrrrrrrrr}
\hline
\multicolumn{4}{c}{\X, $v=0$} && \multicolumn{4}{c}{\ap, $v=0$, $\Omega = 1$ }  \\
\hline
$J$&Obs&Calc&Obs.-Calc. && $J$&Obs&Calc&Obs.-Calc.\\
\hline
   1 &     1.1441 &     1.144 &    0.0001   &&      1   &       2551.297  &    2551.290  &     0.007   \\
   3 &     6.8671 &    6.8641 &    0.0030   &&      2   &       2553.290  &    2553.300  &    -0.010   \\
   5 &    17.1596 &   17.1597 &   -0.0001   &&      3   &       2556.289  &    2556.297  &    -0.008   \\
   7 &    32.0276 &   32.0296 &   -0.0020   &&      4   &       2560.319  &    2560.326  &    -0.006   \\
   9 &    51.4667 &   51.4724 &   -0.0057   &&      5   &       2565.293  &    2565.309  &    -0.015   \\
  11 &    75.4784 &    75.486 &   -0.0076   &&      6   &       2571.361  &    2571.365  &    -0.004   \\
  13 &   104.0584 &  104.0679 &   -0.0095   &&      7   &       2578.322  &    2578.324  &    -0.002   \\
  15 &   137.2076 &  137.2153 &   -0.0077   &&      8   &       2586.413  &    2586.416  &    -0.003   \\
  17 &   174.9176 &  174.9246 &   -0.0070   &&      9   &       2595.338  &    2595.342  &    -0.005   \\
  19 &   217.1868 &  217.1919 &   -0.0051   &&     10   &       2605.470  &    2605.477  &    -0.007   \\
  21 &   264.0096 &  264.0129 &   -0.0033   &&     11   &       2616.352  &    2616.360  &    -0.008   \\
  23 &   315.3829 &  315.3826 &    0.0003   &&     12   &       2628.539  &    2628.545  &    -0.007   \\
  25 &   371.2966 &  371.2959 &    0.0007   &&     13   &       2641.369  &    2641.376  &    -0.006   \\
  27 &   431.7481 &  431.7467 &    0.0014   &&     14   &       2655.606  &    2655.617  &    -0.011   \\
  29 &    496.733 &   496.729 &    0.0040   &&     15   &       2670.374  &    2670.386  &    -0.012   \\
  31 &   566.2387 &  566.2358 &    0.0029   &&     16   &       2686.657  &    2686.689  &    -0.032   \\
  33 &   640.2592 &  640.2601 &   -0.0009   &&     17   &       2703.360  &    2703.387  &    -0.027   \\
  35 &   718.7909 &   718.794 &   -0.0031   &&     18   &       2721.727  &    2721.757  &    -0.030   \\
  37 &    801.828 &  801.8294 &   -0.0014   &&     19   &       2740.350  &    2740.375  &    -0.026   \\
  39 &   889.3578 &  889.3576 &    0.0002   &&     20   &       2760.792  &    2760.815  &    -0.023   \\
  41 &   981.3724 &  981.3695 &    0.0029   &&                                                         \\
  43 &  1077.8619 &  1077.856 &    0.0064   &&                                                         \\
  45 &  1178.8162 &  1178.805 &    0.0108   &&                                                         \\
  47 &  1284.2229 &  1284.209 &    0.0142   &&                                                         \\
  49 &  1394.0739 &  1394.054 &    0.0195   &&                                                         \\
  51 &  1508.3514 &  1508.331 &    0.0205   &&                                                         \\
  53 &  1627.0503 &  1627.026 &    0.0240   &&                                                         \\
  55 &  1750.1546 &  1750.128 &    0.0266   &&                                                         \\
  57 &  1877.6509 &  1877.623 &    0.0279   &&                                                         \\
  59 &  2009.5313 &  2009.498 &    0.0333   &&                                                         \\
  61 &  2145.7752 &  2145.739 &    0.0362   &&                                                         \\
  61 &  2145.7752 &  2145.739 &    0.0362   &&                                                         \\
\hline
\end{tabular}

\end{table}

\begin{table}
\centering
\caption{Selections of the Obs.$-$Calc. residuals, in \cm, for the $A$  and $B$ term values ($v=0$, $+$) for the refined model of $^{24}$Mg$^{16}$O. The Obs. term values are represented by our MARVEL values.}
\label{t:error:A:B}
\begin{tabular}{rrrrrrrrrrrrrrrrrrrrr}
\hline
\multicolumn{4}{c}{\A, $v=0$, $\Omega = 1$, $+$} && \multicolumn{4}{c}{\B, $v=0$, $+$ }  \\
\hline
$J$&Obs&Calc&Obs.-Calc. && $J$&Obs&Calc&Obs.-Calc.\\
\hline
    9   &   3548.585 &   3548.563 &     0.022   &&       0   &    20003.59  &    20003.593 &    0.002   \\
   11   &   3569.710 &   3569.685 &     0.025   &&       2   &    20007.07  &    20007.073 &    0.003   \\
   13   &   3594.843 &   3594.825 &     0.018   &&       4   &    20015.19  &    20015.194 &    0.004   \\
   14   &   3608.936 &   3608.919 &     0.017   &&       6   &    20027.95  &    20027.954 &    0.002   \\
   15   &   3623.999 &   3623.980 &     0.020   &&       8   &    20045.35  &    20045.353 &    0.001   \\
   16   &   3640.197 &   3640.085 &     0.111   &&      10   &    20067.38  &    20067.388 &   -0.007   \\
   17   &   3657.260 &   3657.146 &     0.113   &&      12   &    20094.05  &    20094.057 &   -0.006   \\
   18   &   3675.277 &   3675.263 &     0.014   &&      14   &    20125.34  &    20125.359 &   -0.013   \\
   19   &   3694.343 &   3694.321 &     0.022   &&      16   &    20161.27  &    20161.290 &   -0.011   \\
   20   &   3714.460 &   3714.447 &     0.013   &&      18   &    20201.84  &    20201.845 &   -0.006   \\
   21   &   3735.522 &   3735.500 &     0.021   &&      20   &    20247.01  &    20247.023 &   -0.005   \\
   22   &   3757.646 &   3757.634 &     0.011   &&      22   &    20296.81  &    20296.818 &   -0.001   \\
   23   &   3780.699 &   3780.678 &     0.021   &&      24   &    20351.22  &    20351.224 &    0.004   \\
   24   &   3804.825 &   3804.819 &     0.007   &&      26   &    20410.24  &    20410.238 &    0.002   \\
   25   &   3829.869 &   3829.851 &     0.018   &&      28   &    20473.85  &    20473.854 &    0.001   \\
   26   &   3856.002 &   3855.996 &     0.006   &&      30   &    20542.08  &    20542.065 &    0.019   \\
   27   &   3883.034 &   3883.012 &     0.022   &&      32   &    20614.86  &    20614.864 &    0.000   \\
   28   &   3911.167 &   3911.159 &     0.007   &&      34   &    20692.23  &    20692.245 &   -0.009   \\
   29   &   3940.174 &   3940.157 &     0.018   &&      36   &    20774.18  &    20774.200 &   -0.011   \\
   31   &   4001.292 &   4001.277 &     0.014   &&      38   &    20860.71  &    20860.721 &   -0.007   \\
   32   &   4033.413 &   4033.422 &    -0.009   &&      40   &    20951.81  &    20951.800 &    0.009   \\
        &            &            &             &&      42   &    21047.41  &    21047.428 &   -0.019   \\
\hline
\end{tabular}

\end{table}

\begin{table}
\caption{Selections of Obs. - Calc. residuals, in \cm, showing the vibrational accuracy of the refined model. The Obs. term values are represeted by our MARVEL values. }
\centering
\label{H2D0}
\begin{tabular}{rrrrrrrr}
\hline
$J$&parity&State&$v$&$\Omega$&Obs&Calc&Obs.-Calc.\\
\hline
       1   &      $-$      &      $X$      &           0   &           0   &         1.1441   &       1.144   &      0.0001   \\
       1   &      $-$      &      $X$      &           1   &           0   &       775.8723   &    775.8711   &      0.0012   \\
       6   &       +       &      $X$      &           2   &           0   &      1562.7462   &    1562.715   &      0.0314   \\
       1   &       +       &      $a$      &           0   &           0   &      2551.2967   &     2551.29   &      0.0068   \\
       3   &       +       &      $a$      &           1   &           0   &      3193.0857   &    3193.105   &     -0.0190   \\
       7   &      $-$      &      $A$      &           0   &          -1   &       3539.537   &    3539.509   &      0.0280   \\
       1   &      $-$      &      $A$      &           1   &          -1   &      4160.8965   &    4160.902   &     -0.0058   \\
       3   &       +       &      $A$      &           2   &           1   &      4814.6699   &    4814.617   &      0.0527   \\
       3   &       +       &      $A$      &           3   &           1   &      5455.8189   &    5455.807   &      0.0123   \\
       0   &       +       &      $B$      &           0   &           0   &     20003.5941   &    20003.59   &      0.0016   \\
       0   &       +       &      $B$      &           1   &           0   &     20818.1573   &    20818.14   &      0.0126   \\
\hline
\end{tabular}
\end{table}

\subsection{Partition function}

Partition functions, $Q(T)$, were computed with \Duo\ in steps of 1 K
by explicit summation of the calculated energy levels.  ExoMol always
produces partition functions according to the HITRAN convention
\citep{jt692} which explicitly includes the full atomic nuclear spin
degeneracy, $g_{\rm ns}$. Since the nuclear spins of $^{24}$Mg and
  $^{16}$O are both zero, therefore the nuclear statistical weight is
  1. There is no need to scale the \citet{84SaTaxx.partfunc} or
  \citet{16BaCoxx.partfunc} partition functions for comparison.  The
  nuclear spins of $^{26}$Mg and $^{18}$O are also zero, while the
  $^{25}$Mg and $^{17}$O have the nuclear spins of $5/2$, which result
  in the nuclear spin statistical factor of 6 for $^{25}$Mg$^{16}$O
  and $^{24}$Mg$^{17}$O.

\begin{figure}
\includegraphics[width=0.8\textwidth]{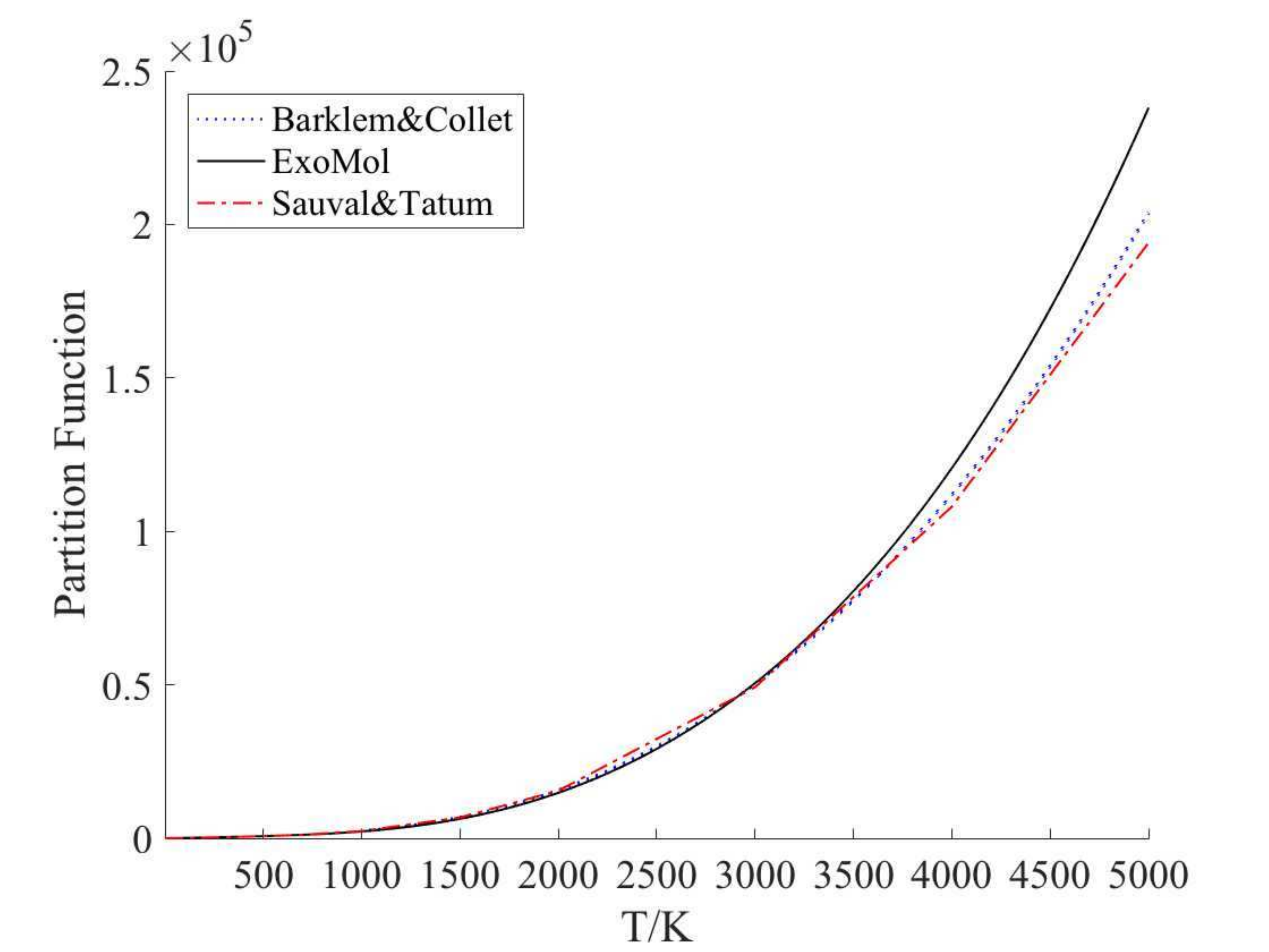}
\hskip-1cm\caption{Partition function of $^{24}$Mg$^{16}$O produced from the ExoMol line list compared with \citet{84SaTaxx.partfunc} and \citet{16BaCoxx.partfunc}. }
\label{fig:pf}
\end{figure}

At 300 K we obtain a value of $Q =374.651$  which is in excellent
agreement with the value of 374.621 used in the JPL database \citep{jpl}.
Figure \ref{fig:pf}
plots our temperature-dependent partition function and compared with
those computed by \citet{84SaTaxx.partfunc} and by
\citet{16BaCoxx.partfunc}. All three data sets agree very well
below 3500~K with the ExoMol results being higher than Barklem \&\ Collet or
Sauval \& Tatum. We believe our energies to be more complete which should explain why our partition function is higher. However, above about 5000~K
electronically excited states of MgO, not considered in this calculation,
will become increasingly thermally occupied. For this reason and because
the abundance of MgO is likely to small in this temperature region
we used a 5000 K upper temperature limit in this study.

The partition functions for all isotopologues are given in the supplementary data.

\subsection{Lifetimes}

Lifetimes for the states are  provided as part of the \Duo\ calculations,
detailed methodology is given by \citet{jt624}.
Lifetimes of the excited states have only been studied experimentally
for the \B\ state and in this case there is disagreement
between
previous studies on the lifetime of the $v=0$ state. The lifetime was
measured to be 32.7$\pm$1.7 ns by \citet{83DiYaDa.MgO}  using
fluorescence decay, 22.5$\pm$1.5 ns by \citet{87BuHeHe.MgO} ($J=1$) using
Stark quantum-beat spectroscopy, and 21.5$\pm$1.8 ns by
\citet{91NaCoMo.MgO} ($J=70$) also using fluorescence decay.
\citet{10MaBeYa.MgO}
calculated  the $J=0$ and $J=70$ lifetimes as 33.3 and 22.0 ns, respectively
suggesting
a significant decrease in lifetime with rotational excitation.

\begin{figure}
\includegraphics[width=0.8\textwidth]{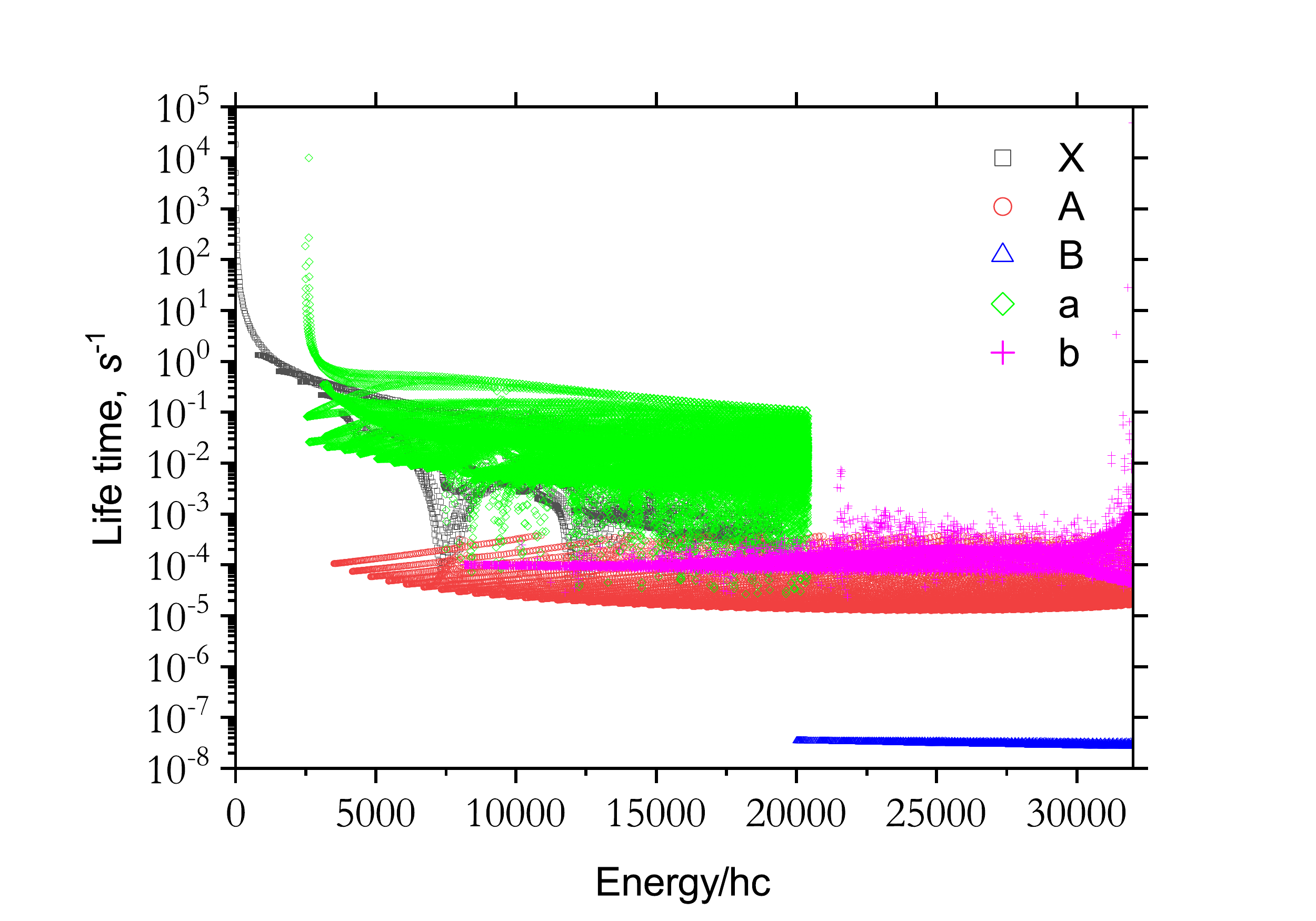}
\caption{Lifetimes of the four low-lying states of $^{24}$Mg$^{16}$O computed by summation of the Einstein-A coefficients.}
\label{fig:lifetime}
\end{figure}

Our lifetimes (\B, $v=0$) for $J=0$ and $J$ =70 are 21.8 ns and 21.7 ns,
respectively, in agreement with the measurements of
\citet{87BuHeHe.MgO} and \citet{91NaCoMo.MgO}.  An overview of the
lifetimes of MgO for the four lowest electronic states is presented in
Figure \ref{fig:lifetime}. The shortest-lived states are from \B. The lowest states ($v=0$, small $J$) of
\X\ and \ap\ have very long lifetime.

\subsection{Line lists}

Line lists were generated by consider all lower states up to
24~000~cm$^{-1}$, upper states up to 37~500~cm$^{-1}$  and rotationally excited
states up to $J = 300$. These parameters are sufficient for completeness
up to temperatures of 5000 K and wavenumbers up to 33~000 ~cm$^{-1}$
or wavelengths longer than 0.3~$\mu$m, although at higher temperatures
the line lists will be not quite complete for the highest wavenumbers.

The line list for $^{24}$Mg$^{16}$O contains 72~833~173 transitions
between 186~842 states. The
line list have been formatted into the ExoMol format (transition files
and state files) \citep{jt548} with lifetimes of each state quoted. A
extract of the states file is shown in Table \ref{table:states} and a
extract of the transition file is shown in Table \ref{trans}. The line lists for isotopologues have comparable sizes.

%25 73881844

\begin{table}
\tt
\centering
\caption{Extract from the states file for $^{24}$Mg$^{16}$O. }
\label{table:states}
\begin{tabular}{rrrrrrcclrrrr}
\hline\hline
$i$&$\tilde{E}$&$g_i$&$J$&$\tau$&$g$&$+/-$&$e/f$&State&$v$&$\Lambda$&$\Sigma$&$\Omega$\\
\hline
     636 &        1.144048 &     3 &    1 &    1.77983E+05 &      0.000000 &    -   &    e   &  X1Sigma+      &     0 &     0 &     0 &     0   \\
     637 &      775.871080 &     3 &    1 &    1.35976E+00 &      0.000000 &    -   &    e   &  X1Sigma+      &     1 &     0 &     0 &     0   \\
     638 &     1540.259721 &     3 &    1 &    6.33385E-01 &      0.000002 &    -   &    e   &  X1Sigma+      &     2 &     0 &     0 &     0   \\
     639 &     2294.147145 &     3 &    1 &    3.98566E-01 &      0.000034 &    -   &    e   &  X1Sigma+      &     3 &     0 &     0 &     0   \\
     640 &     2551.292291 &     3 &    1 &    8.23030E-02 &      0.527645 &    -   &    e   &  a3Pi          &     0 &    -1 &     0 &    -1   \\
     641 &     2622.222505 &     3 &    1 &    2.58488E-02 &     -0.027585 &    -   &    e   &  a3Pi          &     0 &    -1 &     1 &     0   \\
     642 &     3038.170256 &     3 &    1 &    2.22267E-01 &      0.000236 &    -   &    e   &  X1Sigma+      &     4 &     0 &     0 &     0   \\
     643 &     3193.107760 &     3 &    1 &    3.27360E-02 &      0.526013 &    -   &    e   &  a3Pi          &     1 &    -1 &     0 &    -1   \\
     644 &     3265.983987 &     3 &    1 &    2.08686E-02 &     -0.026155 &    -   &    e   &  a3Pi          &     1 &    -1 &     1 &     0   \\
     645 &     3504.293682 &     3 &    1 &    1.05523E-04 &      0.499992 &    -   &    e   &  A1Pi          &     0 &    -1 &     0 &    -1   \\
     646 &     3771.346664 &     3 &    1 &    9.69221E-02 &      0.002761 &    -   &    e   &  X1Sigma+      &     5 &     0 &     0 &     0   \\
     647 &     3824.775932 &     3 &    1 &    2.06126E-02 &      0.520372 &    -   &    e   &  a3Pi          &     2 &    -1 &     0 &    -1   \\
     648 &     3901.850493 &     3 &    1 &    1.82913E-02 &     -0.023039 &    -   &    e   &  a3Pi          &     2 &    -1 &     1 &     0   \\
     649 &     4160.902293 &     3 &    1 &    7.47291E-05 &      0.499992 &    -   &    e   &  A1Pi          &     1 &    -1 &     0 &    -1   \\
     650 &     4446.228208 &     3 &    1 &    1.51760E-02 &      0.538037 &    -   &    e   &  a3Pi          &     3 &    -1 &     0 &    -1   \\
     651 &     4479.741149 &     3 &    1 &    2.31664E-02 &     -0.028181 &    -   &    e   &  a3Pi          &     3 &    -1 &     1 &     0   \\
     652 &     4544.158030 &     3 &    1 &    2.77669E-02 &     -0.009762 &    -   &    e   &  X1Sigma+      &     6 &     0 &     0 &     0   \\
     653 &     4809.732862 &     3 &    1 &    5.84090E-05 &      0.499993 &    -   &    e   &  A1Pi          &     2 &    -1 &     0 &    -1   \\
     654 &     5057.431336 &     3 &    1 &    1.21026E-02 &      0.530620 &    -   &    e   &  a3Pi          &     4 &    -1 &     0 &    -1   \\
     655 &     5117.344124 &     3 &    1 &    1.20824E-02 &     -0.030073 &    -   &    e   &  a3Pi          &     4 &    -1 &     1 &     0   \\
     656 &     5239.189020 &     3 &    1 &    7.80934E-02 &     -0.000452 &    -   &    e   &  X1Sigma+      &     7 &     0 &     0 &     0   \\
     657 &     5450.913256 &     3 &    1 &    4.83072E-05 &      0.499994 &    -   &    e   &  A1Pi          &     3 &    -1 &     0 &    -1   \\
     658 &     5658.311999 &     3 &    1 &    1.01359E-02 &      0.528937 &    -   &    e   &  a3Pi          &     5 &    -1 &     0 &    -1   \\

\hline\hline
\end{tabular}
\mbox{}\\

{\flushleft
$i$:   State counting number.     \\
$\tilde{E}$: State energy in \cm. \\
$g_i$:  Total statistical weight, equal to ${g_{\rm ns}(2J + 1)}$.     \\
$J$: Total angular momentum.\\
$\tau$: Lifetime (s$^{-1}$).\\
$g$: Land\'{e} $g$-factors. \\
$+/-$:   Total parity. \\
$e/f$:   Rotationless parity. \\
State: Electronic state.\\
$v$:   State vibrational quantum number. \\
$\Lambda$:  Projection of the electronic angular momentum. \\
$\Sigma$:   Projection of the electronic spin. \\
$\Omega$:   Projection of the total angular momentum, $\Omega=\Lambda+\Sigma$. \\
}
\end{table}

\begin{table}
\tt
\caption{Extract from the transitions file for $^{24}$Mg$^{16}$O. }
\centering
\begin{tabular}{lllr}
\hline
$f$&$i$&$A_{fi}$(s$^{-1}$)&$\tilde{\nu}$ (cm$^{-1}$)\\
\hline
       36757   &    37183  &  7.7028E-12      &        0.098041    \\
       32237   &    31809  &  2.1383E-14      &        0.098067    \\
       32909   &    32497  &  1.0981E-11      &        0.098167    \\
       90382   &    90758  &  2.8824E-13      &        0.098189    \\
       98009   &    98376  &  3.5386E-14      &        0.098192    \\
       49202   &    49606  &  1.5586E-12      &        0.098197    \\
      123918   &   124241  &  1.7154E-15      &        0.098208    \\
       91640   &    91262  &  4.8358E-13      &        0.098261    \\
       83152   &    82769  &  9.9825E-15      &        0.098285    \\
        7957   &     8392  &  4.4448E-11      &        0.098298    \\
      101679   &   102046  &  5.2911E-14      &        0.098322    \\
       82876   &    83259  &  2.4564E-18      &        0.098324    \\
\hline
\end{tabular}

\noindent
\mbox{}\\
{\flushleft
$f$=state number of final state\\
$i$=state number of inital state\\
$A_{fi}$=Einstein-A coefficient\\
$\tilde{\nu}$=transition wavenumber\\
}
\label{trans}
\end{table}

\section{Simulated spectra}

\begin{figure}
\includegraphics[width=0.8\textwidth]{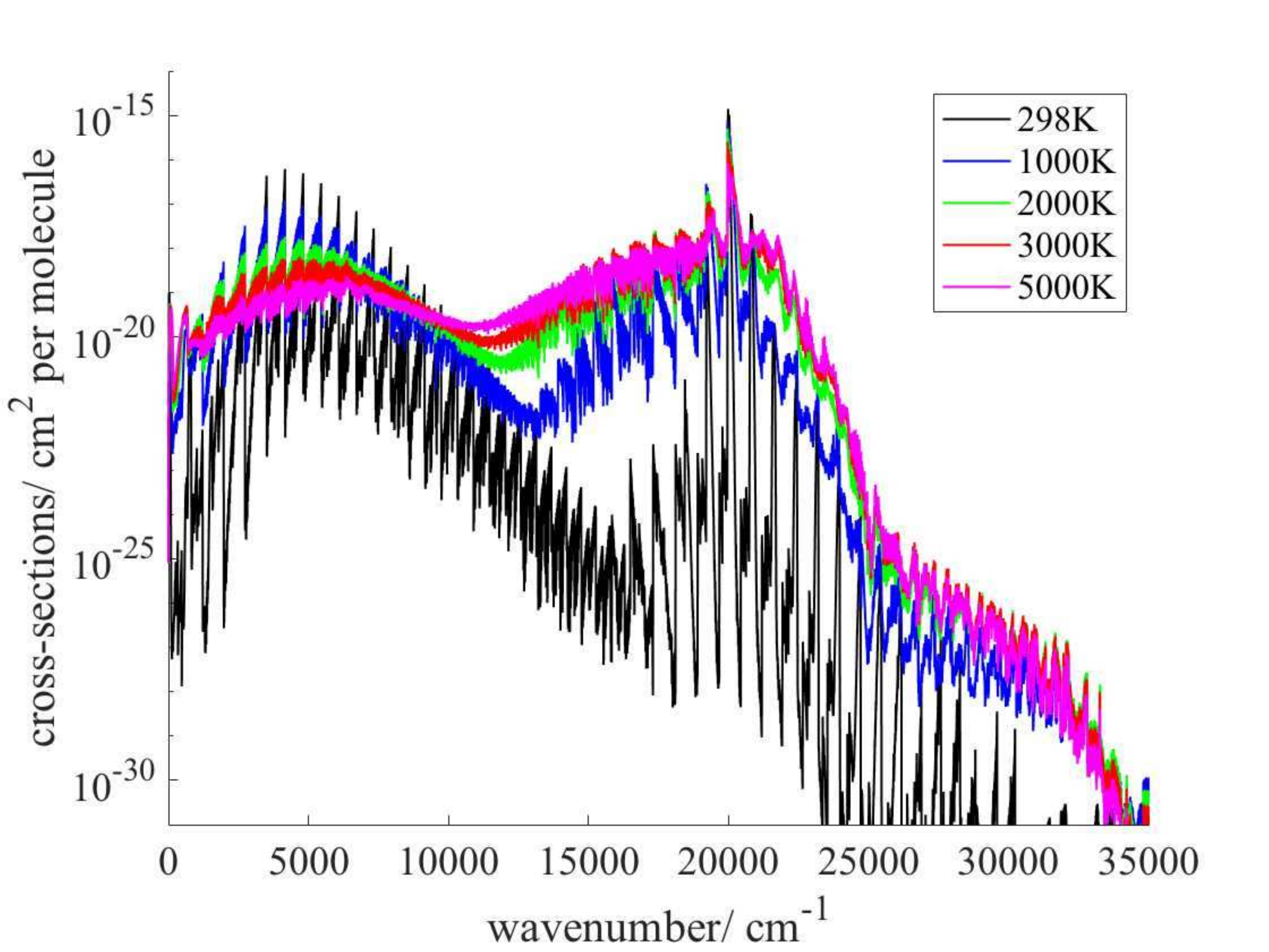}
\caption{Simulated absorption spectra of MgO at 5 different temperatures.
The spectrum becomes increasingly flat with increasing temperature:  the difference with temperature is most marked around 20000 cm$^{-1}$. }
\label{fig:absop_temp}
\end{figure}

\begin{figure}
\includegraphics[width=0.8\textwidth]{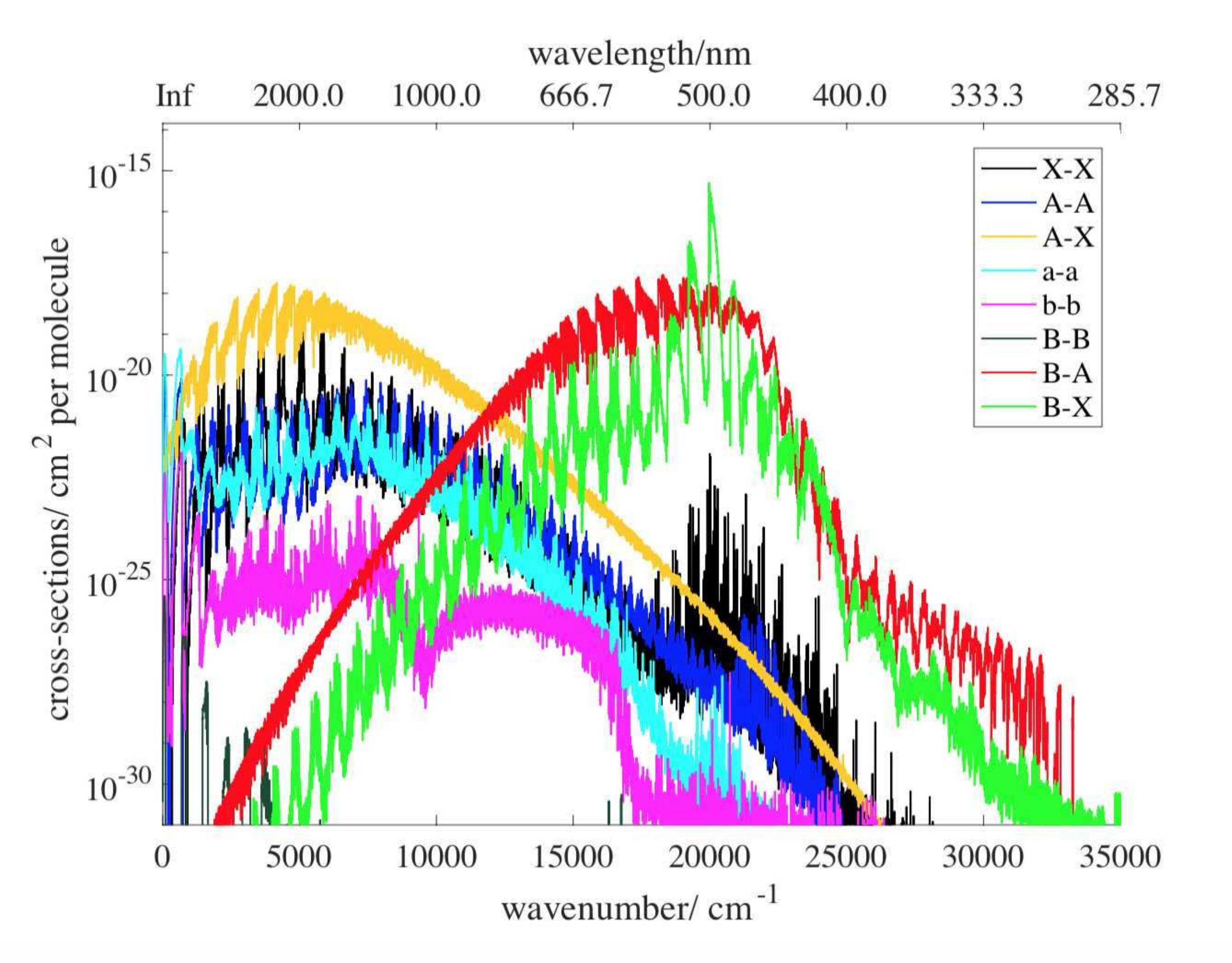}
\caption{Contribution of each band to the simulated absorption spectra of MgO at 2000 K. }
\label{fig:absop_states}
\end{figure}

Absorption spectra at different temperatures are presented in Figure
\ref{fig:absop_temp}. The contribution of each band is shown in Figure
\ref{fig:absop_states}. The green ($B-X$) and red ($B-A$) singlet band systems
are the strongest bands and show a significant overlap at visible
wavelengths. All spectral simulations employed the \xcross\ code~\citep{jt708}.

\begin{figure}
\hskip-1cm\includegraphics[width=0.8\textwidth]{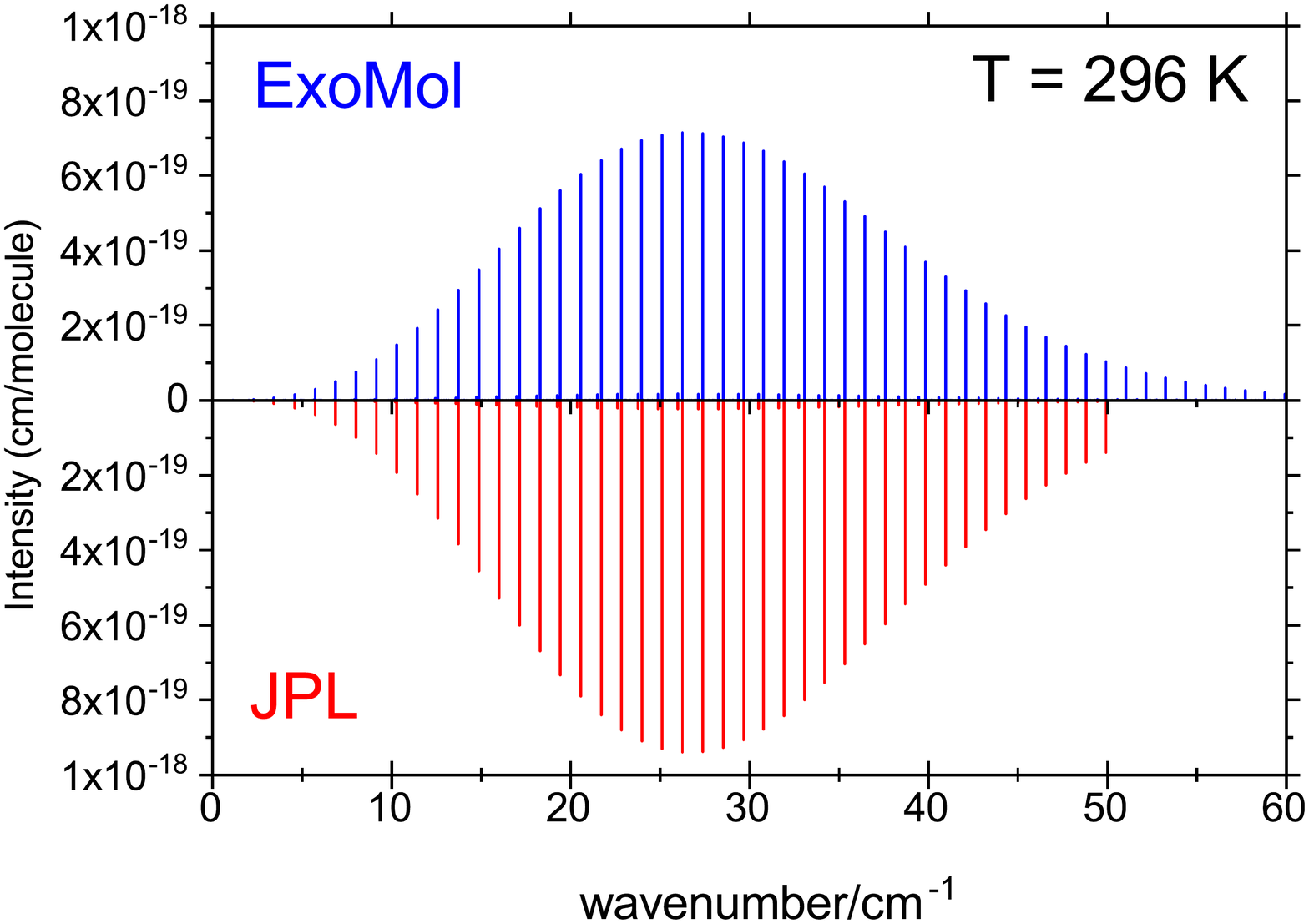}
\caption{Comparison of the ExoMol spectrum and JPL. }
\label{fig:JPL}
\end{figure}

Comparisons with the previous experimental works have been made to
assess the quality of our computed line list.  The rotational $X$ band in the form of a stick spectrum is shown in Figure~\ref{fig:JPL},
where it is compared to the experimentally determined transitions from the JPL database \citep{jpl}.
The JPL spectrum is based on the dipole moment of 6.88 D  taken from an old \ai\ calculation due
to
\citet{91FoSaxx.MgO}.
Our vibrational averaged ($v=0$) value is 5.99~D, which is the same as given by \citet{17BaScxx.MgO}.
We expect this new value to be represent an improvement which suggests that the current
JPL intensities are about 30\%\ too strong and, correspondingly, the observational upper limits
for the  ISM \citep{85TuStxx.MgO,98SaWhKa.MgO} to be about 30\%\ too low.

\subsection{The red band}

\begin{figure}
\hskip-1cm\includegraphics[width=0.8\textwidth]{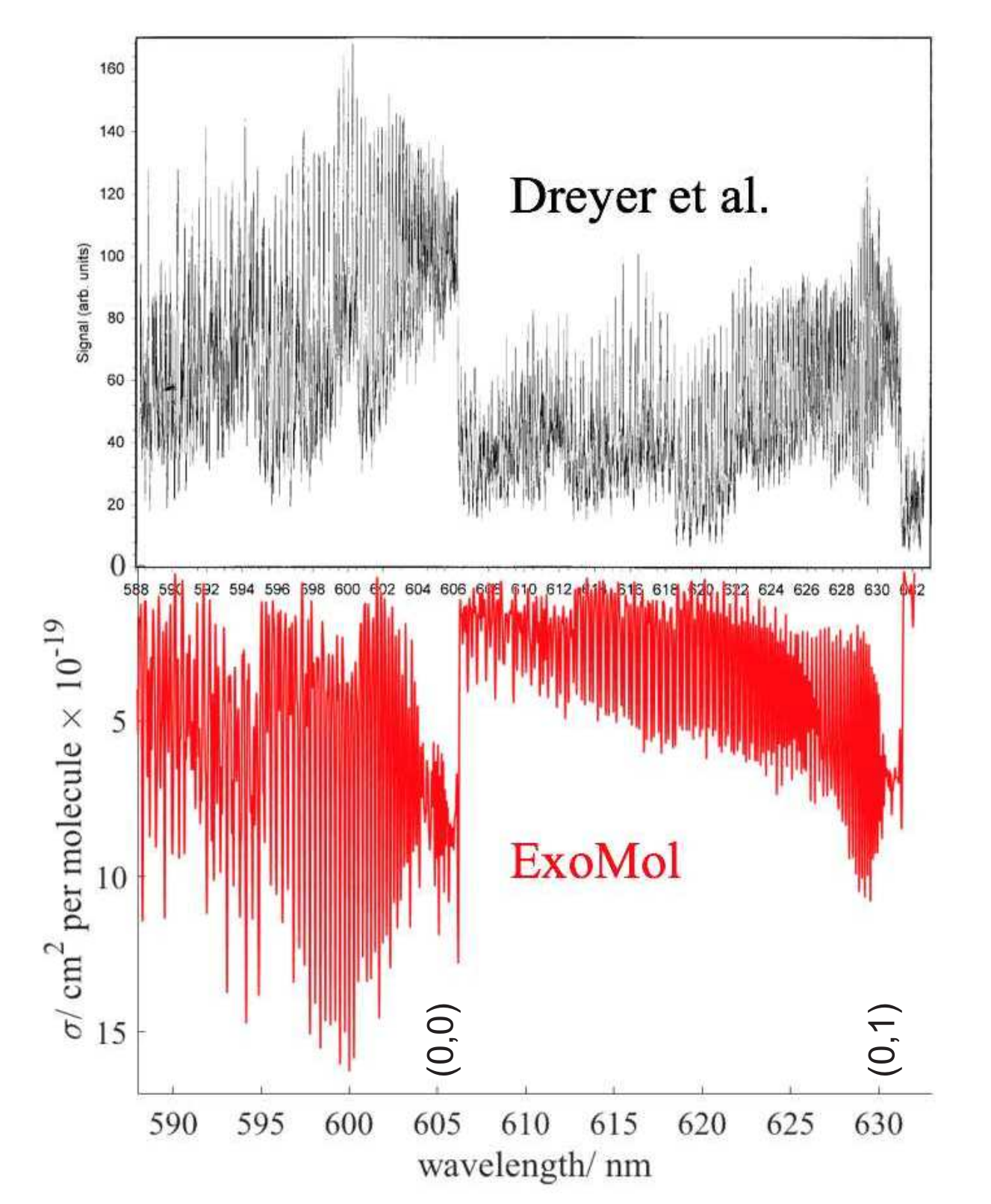}
\caption{Comparison of the ExoMol spectrum (lower) and experimental spectrum from \citet{01DrDaAb.MgO} (upper) in the range of 588 nm to 633 nm. The (0,0) and (0,1) bandheads agree whereas the other bandheads cannot be clearly distinguished from each other.  }
\label{fig:Dreyer1}
\end{figure}

\citet{01DrDaAb.MgO} made excitation scans using the Laser-induced
fluorescence excitation spectroscopy for the $B-A$ transitions.  For
the 588 nm to 632 nm range there is overall agreement as shown in Figure
\ref{fig:Dreyer1}. The $(0,0)$ and $(0,1)$ bandheads give a
good match, whereas the bandheads for the higher state are not as clear as
the lower ones.

\begin{figure}
\hskip-1cm\includegraphics[width=0.90\textwidth]{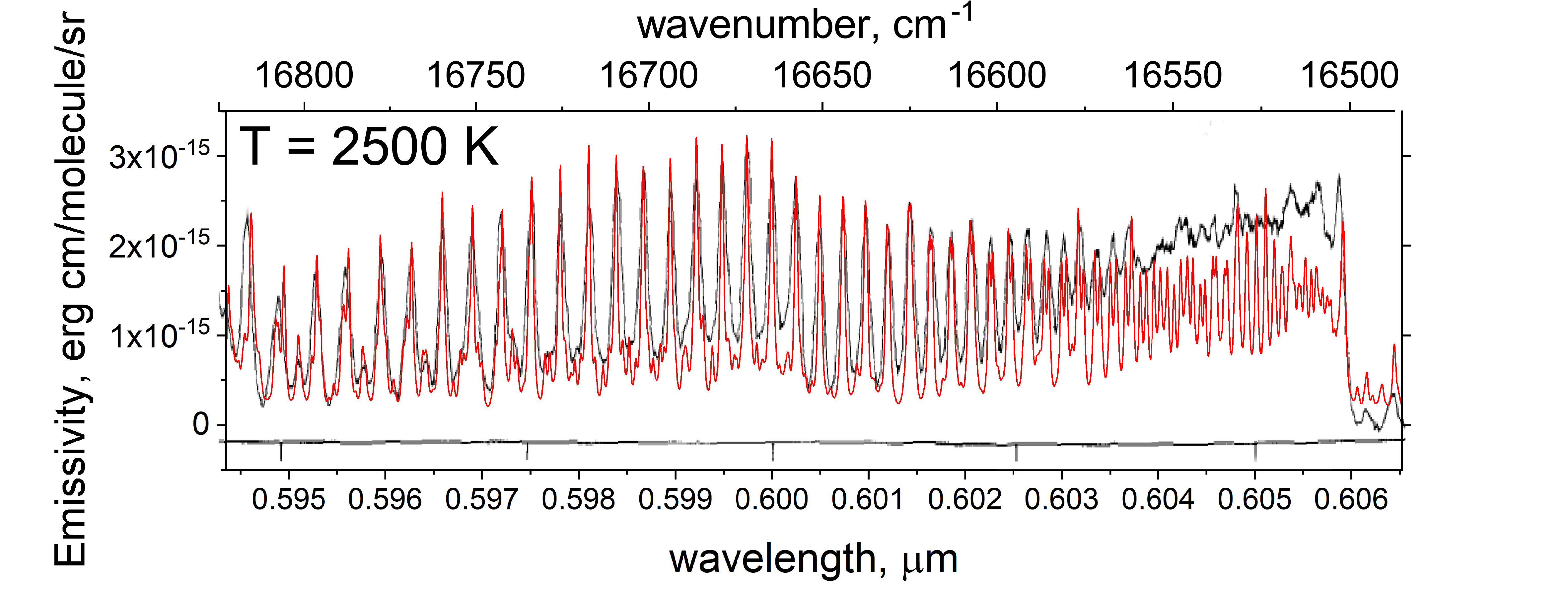}
\caption{Comparison of the ExoMol emission (2300~K) and experimental spectra  from  \citet{01DrDaAb.MgO} in the range of 16115 cm$^{-1}$ to 16150 cm $^{-1}$ using the Voigt profile with $\gamma=1.1$~\cm\ combined from the contributions from three main isotopologues, also shown, assuming the terrestrial abundance (0.79/0.10/0.11 for 24/25/26, respectively).}
\label{fig:Dreyer2}
\end{figure}

\begin{figure}
\hskip-1cm\includegraphics[width=0.8\textwidth]{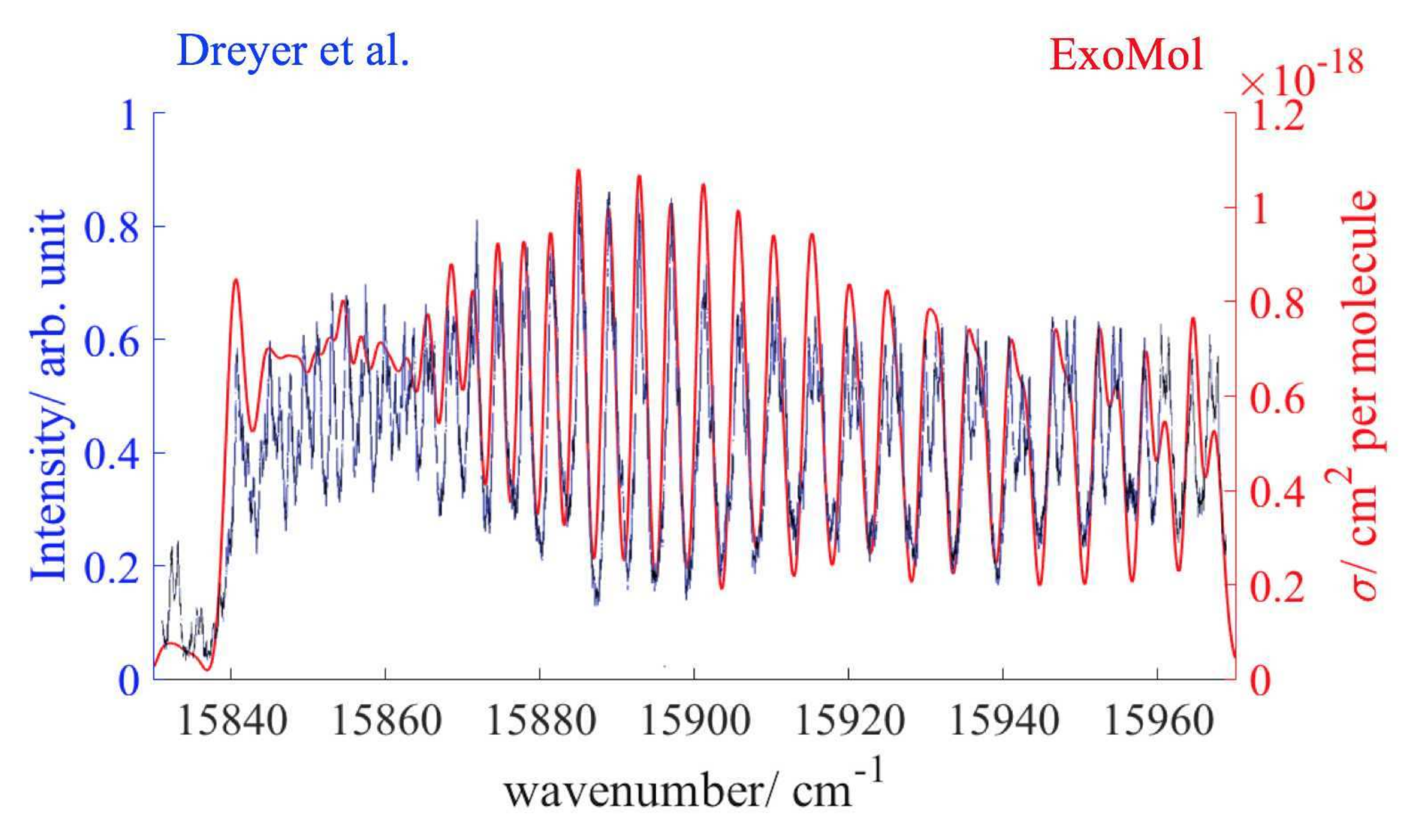}
\caption{Comparison of the ExoMol absorption spectrum and experimental spectrum from \citet{01DrDaAb.MgO} in the range of 15840 cm$^{-1}$ to 15960 cm $^{-1}$ for the (1,0) band using the Gaussian line profile of HWHM=1~\cm\ at $T= 2300$~K.}
\label{fig:Dreyer3}
\end{figure}

Figure \ref{fig:Dreyer2} shows a spectrum for the 16115 cm$^{-1}$ to 16150 cm $^{-1}$  region from \citet{01DrDaAb.MgO}.
The main peaks for the $(0,1)$ lines generally agree whereas the
magnitudes can differ, see Figure \ref{fig:Dreyer3}. The $(0,1)$
bandhead region cannot be resolved between 1584 nm and 1586 nm
otherwise the resolved peaks agree with experiment within 0.1 nm.
This figure shows the individual contributions from all three major isotopologues assuming the terrestrial abundance (0.79/0.10/0.11 for $^{24}$Mg$^{16}$O/$^{25}$Mg$^{16}$O/$^{26}$Mg$^{16}$O, respectively).

\begin{figure}
\hskip-1cm\includegraphics[width=1.0\textwidth]{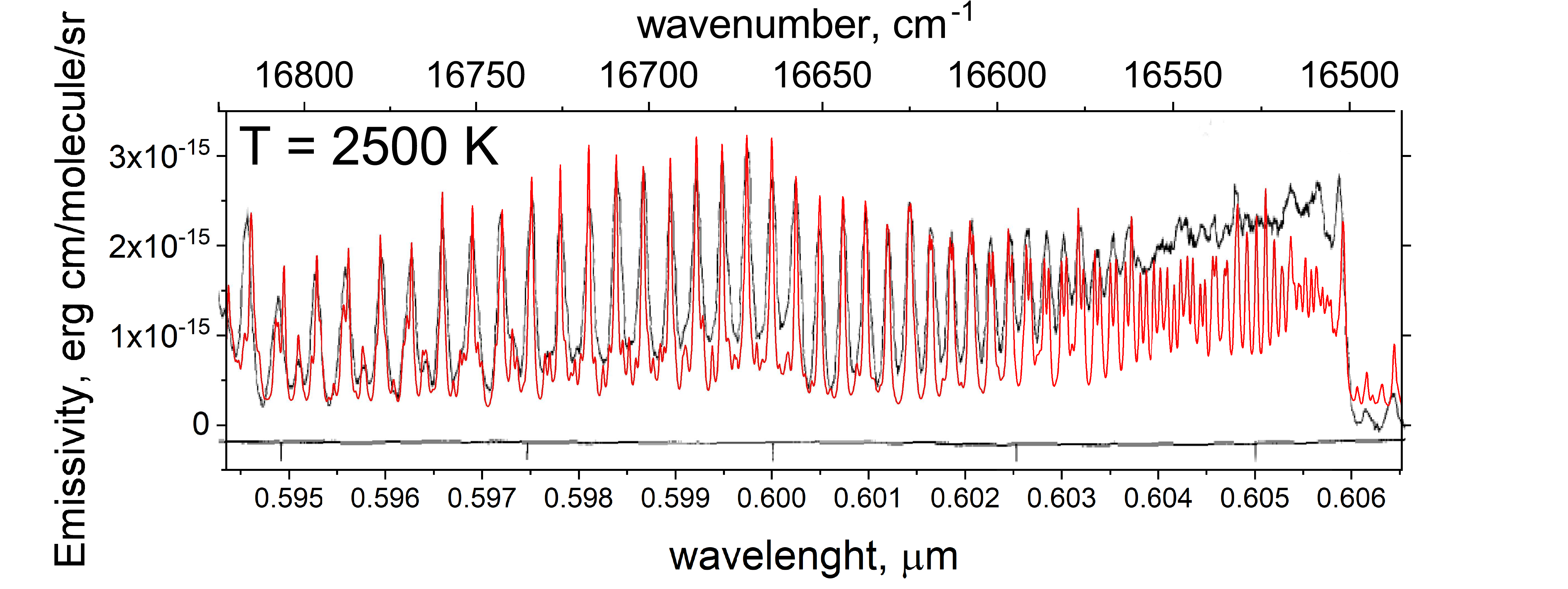}
\caption{Comparison of the ExoMol emission spectrum and experimental spectrum from \citet{78PaBaMc.MgO} around the (1,1) bandhead using the Voigt profile with $\gamma=1.1$~\cm\ at $T= 2500$~K. The vacuum wavelength was re-calibrated to the acetylene-air wavelength to match the experiment. }

\label{fig:Pasternack}
\end{figure}

Red band studies made by \citet{78PaBaMc.MgO} also show good agreement
with the ExoMol data around the $(1,1)$ band; Figure
\ref{fig:Pasternack} gives a comparison of the emission
spectrum simulated using the ExoMol line list with the laser-induced
fluorescence of MgO in acetylene-air flames \citep{78PaBaMc.MgO}. In
order to match the experiment, the vacuum wavelength was re-scaled to
the air-acetylene wavelength using the refractive indexes from
\citep{96Ciddor,09Loria}\footnote{The refractive index data are taken
  from M. N. Polyanskiy's ``Refractive index database'',
  https://refractiveindex.info.} assuming a 45:55 mixture. This
ratio was adjusted to match the experimental spectrum.

\subsubsection{The green band}

\begin{figure}
\includegraphics[width=0.8\textwidth]{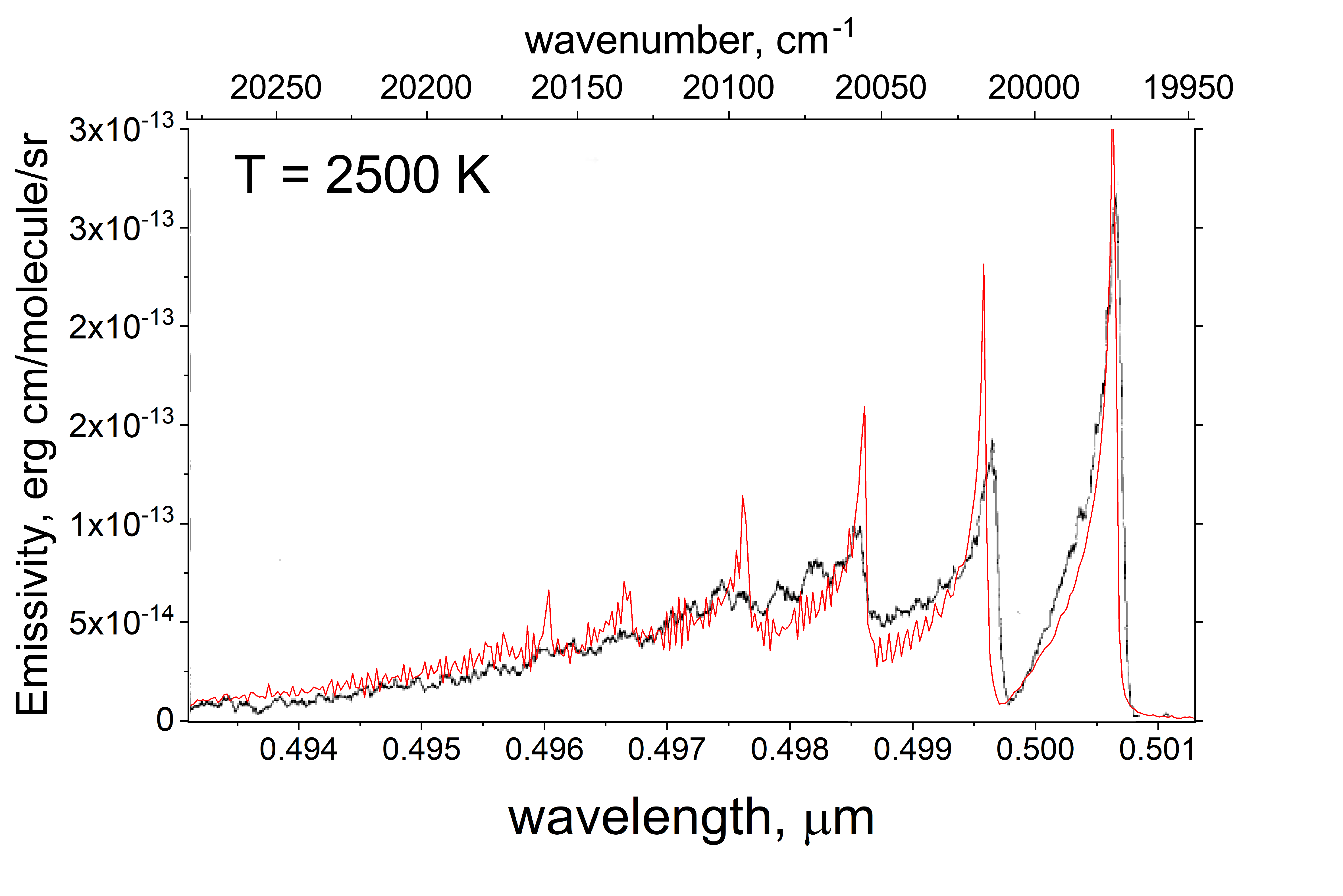}
\caption{Comparison of the Exomol emission spectrum and experimental spectrum from \citet{78PaBaMc.MgO} in the range of 495 nm to 501 nm. The ExoMol spectrum was generated using the Voigt profile with $\gamma=1.1$~\cm\ at $T= 2500$~K. The vacuum wavelength was re-calibrated to the acetylene-air wavelength to match the experiment.  }
\label{fig:Pasternack2}
\end{figure}

\begin{figure}
\includegraphics[width=0.8\textwidth]{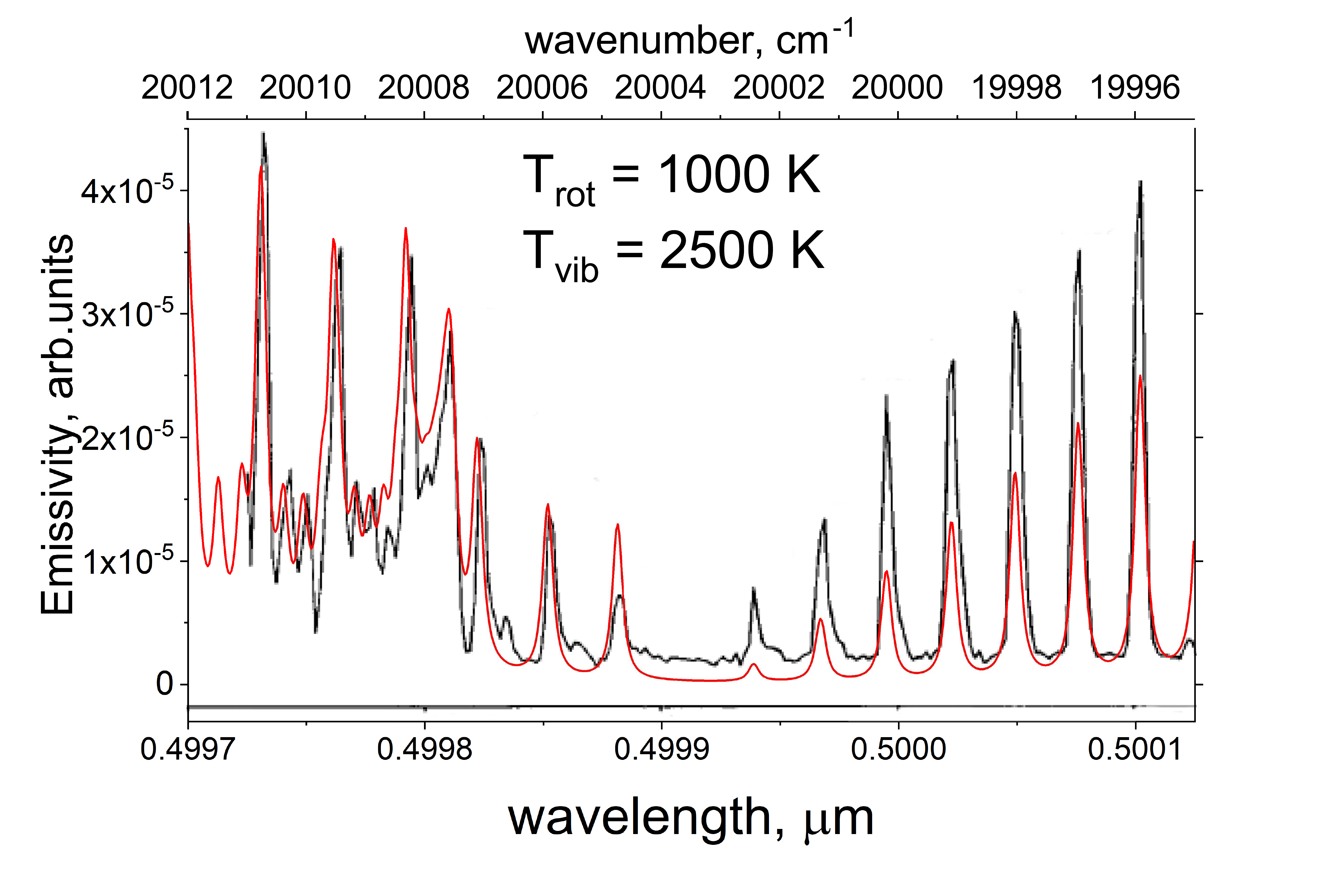}
\caption{Comparison of the absorption ExoMol spectrum and experimental spectrum from \citet{87BuHeHe.MgO} in the range of 499.7 nm to 500.1 nm in the green band using the Gaussian line profile of HWHM=1~\cm, at $T_{\rm rot}= 1000$~K and $T_{\rm rot}= 2500$~K. }
\label{fig:Brusener}
\end{figure}

For the 495 nm to 501 nm region of the $B-X$ transitions,
Figure~\ref{fig:Pasternack2} compares our data with the laser-induced fluorescence spectrum of
\citet{78PaBaMc.MgO}. The shapes, which are largely determined by bandheads, of the  two spectra are similar.
The same wavelength calibration (vacuum to air-acetylene) as above was applied. The ExoMol spectrum agrees very well also with the higher resolution experimental data from the same region, measured between 499.7 nm and 500.1 nm using Stark quantum-beat spectroscopy by \citet{87BuHeHe.MgO}, see Figure \ref{fig:Brusener}, where no calibration was required. For these simulations the non-LTE model was assumed based on two temperatures, rotational $T_{\rm rot}$ and vibrational $T_{\rm vib}$, as implemented in \Duo. The temperatures selected to match the experimental emission spectra ere 1000~K and 2500~K, respectively.

%\red{Re-do this}
%This suggests that Dagdigian's spectrum may not be correctly calibrated.

\section{Conclusion}

Accurate and complete line lists, known as LiTY, for $^{24}$Mg$^{16}$O,
$^{25}$Mg$^{16}$O, $^{26}$Mg$^{16}$O are computed using high level
\ai\ DMCs as well as a spectroscopically refined \Duo\ model. Our work
displays good agreement with the most theoretical and experimental
studies made in the past.  The lifetimes computed for
($B\,{}^1\Sigma^+$, $v=0$) are 2.18 ns for $J=0$ and 2.17 ns for
$J=70$, partially matching prior studies which disagree with each
other. The partition functions agree with other theoretical studies
for the temperature up to 5000 K

With a valid temperature range up to 5000 K, the strong red and green
bands between 15000 and 20000 cm$^{-1}$ (0.66 to 0.5 $\mu$m) should be
useful for detecting MgO in the atmospheres of cool stars, brown
dwarfs and some exoplanets.

The LiTY line lists can be downloaded from the CDS, via
ftp://cdsarc.u-strasbg.fr/pub/cats/J/MNRAS/, or
http://cdsarc.u-strasbg.fr/viz-bin/qcat?J/MNRAS/, or from
www.exomol.com.

\section*{Acknowledgements}

This work was supported by the UK Science and Technology Research
Council (STFC) No. ST/R000476/1 and the COST action MOLIM No. CM1405.
This work made extensive use of UCL's Legion high performance
computing facility  along with the STFC DiRAC HPC facility
supported by BIS National E-infrastructure capital grant ST/J005673/1
and STFC grants ST/H008586/1 and ST/K00333X/1.
We thank Charles Bauschlicher for providing their \ai\ PECs and DMCs of MgO.

\bibliographystyle{mn2e}
%\bibliography{journals_astro,MgO,air,C2H2,CaO,jtj,methods,linelists,a-models,MARVEL,partition,programs,additional,exoplanets,exogen,TiO}

\label{lastpage}

\end{document}